\documentclass[referee]{raa}
\usepackage{graphicx,times}
\usepackage{amssymb}
\begin{document}

   \title{Modeling blue stragglers in young clusters}

   \volnopage{Vol.0 (200x) No.0, 000--000}
   \setcounter{page}{1}

   \author{Pin Lu
      \inst{1,2}
   \and Li-Cai Deng
      \inst{1}
   \and Xiao-Bin Zhang
      \inst{1}
   }
   \institute{Key Laboratory of Optical Astronomy, National Astronomical Observatories, Chinese Academy of Sciences,
             Beijing 100012, China; {\it lupin@bao.ac.cn}\\
        \and
             Graduate University of Chinese Academy of Sciences, Beijing, 100049, China\\
   }

   \date{Received~~2011 month day; accepted~~2011~~month day}

\abstract{In this paper, a grid of the binary evolution models are
calculated for the study of blue straggler (BS) population in
intermediate age ($\log Age$=7.85-8.95) star clusters. The BS
formation via mass transfer and merging is studied systematically
using our models. Both Case A and B close binary evolutionary tracks
are calculated in a large range of parameters. The results show that
BSs formed via Case B are generally bluer and even more luminous
than those produced by Case A. Furthermore, the larger range in
orbital separations of Case B models provide a probability of
producing more BSs than Case A. Based on the grid of models, several
Monte-Carlo simulations of BS populations in the clusters in the age
range are carried out. The results show that BSs formed via
different channels populate different areas in color magnitude
diagram(CMD). The locations of BSs in CMD for a number of clusters
are compared to our simulations as well. In order to investigate the
influence of mass transfer efficiency in the models and simulations,
a set of models are also calculated by implementing a constant mass
transfer efficiency, $\beta$=0.5 during Roche lobe overflow (Case A
binary evolution excluded). The result shows BSs can be formed via
mass transfer at any given age in both cases. However, the
distributions of the BS populations on CMD are different.
\keywords{stars: blue stragglers
--- stars: binaries: close
} }

   \authorrunning{Pin Lu, Li-Cai Deng \& Xiao-Bin Zhang }
   \titlerunning{Modeling blue stragglers in young clusters }

   \maketitle

\section{Introduction}
\label{sect:intro}

Blue stragglers are observed in almost all stellar systems which
appear to be anomalously young compared to other populations in the
system. They are commonly defined as stars brighter and bluer than
the turn-off(TO) point on CMD. Such stars seem to remain on the main
sequence(MS) longer than is estimated by standard theory of stellar
evolution. The existence of BSs can significantly affect the
integrated spectral energy distributions, especially  in blue and UV
bands (Li \& Han 2009, Xin, Deng \& Han 2007), thus challenges
traditional single stellar population(SSP).

Ferraro et al.(1993) found two groups of BSs in M3, which have a
bimodal radial distribution. They concluded whether the two groups
of BSs have different physical properties or they represent the
products of different formation mechanisms. Laget et al.(1994)
identified 24 BSs in a UV imaging study of M3. They found two
distinct classes of BSs in M3. One has normal UV flux distribution
and may be the massive stars formed through stellar collisions,
whereas the other has UV excess and appears to be composite systems
consisting of a hot secondary star orbiting a MS turn-off star or a
subgiant or even possibly a low luminosity giant. Recently, Ferraro
et al.(2009) observed two distinct sequences of BS populations on
the CMD of M30. They concluded that the bluer sequence is arising
from direct collisions and the redder one arising from the evolution
of close binaries possibly still experiencing mass exchange. All the
observations above suggest that BSs may have different formation
mechanisms.

Though BS formation mechanisms are not completely understood yet,
many possible scenarios now have been proposed to explain BS
formation (see the review of Stryker 1993). At present, some
possible ways are believed to account for the BS formation: mass
transfer (MT) or complete coalescence in close binaries; stellar
mergers resulting from direct collisions. It is believed that more
than one scenarios are necessary to explain the observations.
However, the relative importance of the two possible mechanisms
remains unclear.

The mechanism of mass transfer in close binary was first proposed by
McCrea (1964). Mass transfer can occur when the massive component in
the binary system expands out of the Roche lobe. As a result of mass
transfer, the companion can become a more massive main-sequence star
than turn-off stars whose life time could be notably extended, or
even doubled compared to a star with the same final mass.
Coalescence of the two components may also occur during the process.
Three mass transfer cases (case A, case B and case C) are defined
according to the evolutionary state of the primary at the onset of
mass transfer. They are associated with the evolutionary stages
respectively: case A on the main sequence, case B after the main
sequence before helium ignition and case C during central He-burning
and afterwards (Kippenhahn \& Weigert 1967). The mass of BS formed
this way cannot exceed twice the cluster turn-off mass. Besides blue
stragglers, many other observational objects are also associated
with interactive binaries, such as Ba stars which can be explained
by mass transfer through wind accretion, disk accretion or common
envelope ejection (Han et al. 1995; Liu et al. 2009). Binary
interactions are sometimes linked to variable stars observed in star
clusters. It is known that some cluster variable stars are still in
binary systems, such as symbiotic stars, W UMa stars and EA-type
binaries. A series of work searching for variables in clusters have
been performed for now (Lu et al. 2006,2007; Xin et al. 2002; Zhang
et al. 2003,2005; Li et al. 2004). Moreover, binaries (white dwarf +
MS or red giant) are believed to be possible progenitors of type Ia
supernova (Meng et al. 2006,2009; Wang et al. 2008,2010; Guo et al.
2008).

BSs could also, in principle, be formed in direct collisions between
main sequence stars. Direct collision hypothesis was originally
presented by Hills \& Day (1976). They proposed that the remnant of
two main-sequence stars collision could also produce a blue
straggler. It is likely the cause of making BSs in globular
clusters, where star densities are extremely high and dynamical
effects are thought to play a dominant role. However, it is unlikely
to happen for open clusters, as the time-scale of collision is too
long (Press \& Teukolsky 1977; Mardling \& Aarseth 2001) and neither
the peculiar compositions of Ap and Bp-type blue stragglers, nor
different rotation rates of observed BSs are by themselves sufficient to explain the
large color differences between observed BSs and MS stars (Mermilliod 1982).

Different mechanisms are believed to dominant in different
environments. As a consequence of dynamic evolution of the host
cluster, the collision scenario is thought to be responsible for BSs
in dense cluster cores. Binary interaction process and products are
believed to dominate in more spare environment such as open clusters
and in the field (Mapelli et al. 2004; Lanzoni et al. 2007;
Dalessandro et al. 2008; Sollima et al. 2008).

Many detailed work about the two mechanisms is presented. Pols \&
Marinus (1994) performed Monte-Carlo simulations of close binary
evolution in young clusters. The BSs predicted by their work agree
well with the observations in quantitatively for clusters younger
than about 300 Myr but are not enough for the clusters of older ages
(between 300 and 1500 Myr). N-body simulations of BSs in old open
cluster M67 (Hurly et al. 2001,2005) suggest that the formation of
BSs is dominated by both mass transfer and cluster dynamics.

Recently, the binary explanation was claimed to be dominant even in
dense globular cluster cores (Knigge \& Sills 2009). They found a
clear, but sub-linear correlation between the number of BSs in a
cluster core and the total stellar mass contained within it. It is
regarded as the strongest and most direct evidence to date that most
blue stragglers, even those found in cluster cores, are the
progeny of binary systems.

Some other mechanisms can also contribute to BS population. BSs
predicted in mass transfer theory are fainter than 2.5 mag above the
cluster turn-off stars, but there are exceptions in observations
such as F81 in M67. To explain these massive stragglers, Hagai and
Daniel (2009) discussed the possibility for BS formation in
primordial and/or dynamical hierarchical triple stars. Chen \& Han
(2008) estimated the remnants of binary coalescence from case A
models. The outcome BSs in their work can also be able to predict
some high luminous BSs, however, the small quantity can hardly match
the luminous BSs in M67. Angular momentum loss (AML) in low mass
binaries can also contribute to BS populations, but only in old
clusters. There are a number of subjects including the treatment of
AML (Li et al. 2004; Demircan et al. 2006; Micheal \& Kevin 2006;
Stepien 2006). The recent study of Chen \& Han (2008a,2009) also
demonstrated that AML is likely a main factor to BS formation in old
open clusters. However, its contribution can be ignored in clusters
younger than 1.0 Gyr.

It is more difficult to study BSs in young clusters than in old ones
in a systematic manner, as the observed BSs in younger clusters are
usually very limited due to the relative small number of massive
main sequence stars. We investigated case A mass transfer in close
binaries in young clusters NGC 1831, whose log(age) is around 8.65
by isochrone fitting. The result shows that binaries experiencing
case A mass transfer can hardly account for the luminous BSs in such
young clusters (Lu \& Deng 2008). A subsequent work combined case A
and B is then performed to study BSs in old open clusters. The
evolutionary behaviors of case A and B are compared in detail. We
concluded that case B is as important as case A in old clusters, and
it is even more important in younger clusters by comparing the
different evolutionary behaviors with different initial primary
mass. We predicted that BSs formed via case B mass transfer can be
brighter and bluer than case A, and the number of BSs can be
remarkably increased (Lu et al. 2010, hereafter LDZ10) thus provide
a clue to study the statistics of BSs in such young clusters.

In this work, we apply the same approach as in LDZ10 to
systematically study BS populations in younger stellar populations
represented by open clusters. We select clusters younger than 1.0
Gyr in this paper, because stellar densities in young clusters are
sufficiently low that dynamical effects are not so important to
complicate our study. Furthermore, there is good evidence for the
presence of a significant fraction of primordial binaries in such
young clusters, making it not only interesting but also essential to
better understand the evolution of the cluster. A large grid of
binary evolutionary calculations with the initial mass of the donor
from 2.0-7.6 $M_\odot$ are carried out in this work. The models can
cover all close binaries experiencing mass transfer between 70.0 Myr
and1.0 Gyr.

We describe and analyse our models in Sect 2. Several Monte-Carlo
simulations are presented and the results are shown in Sect 3. We
will discuss the specific frequency of $N_{\rm BS}$/$N_2$ of our
simulations and compare our results to the observations in Sect 4.
Summaries and conclusions are given in the final section.

\section{The binary models}
\label{sect:model}

\begin{figure}
\includegraphics[width=45mm,height=75mm]{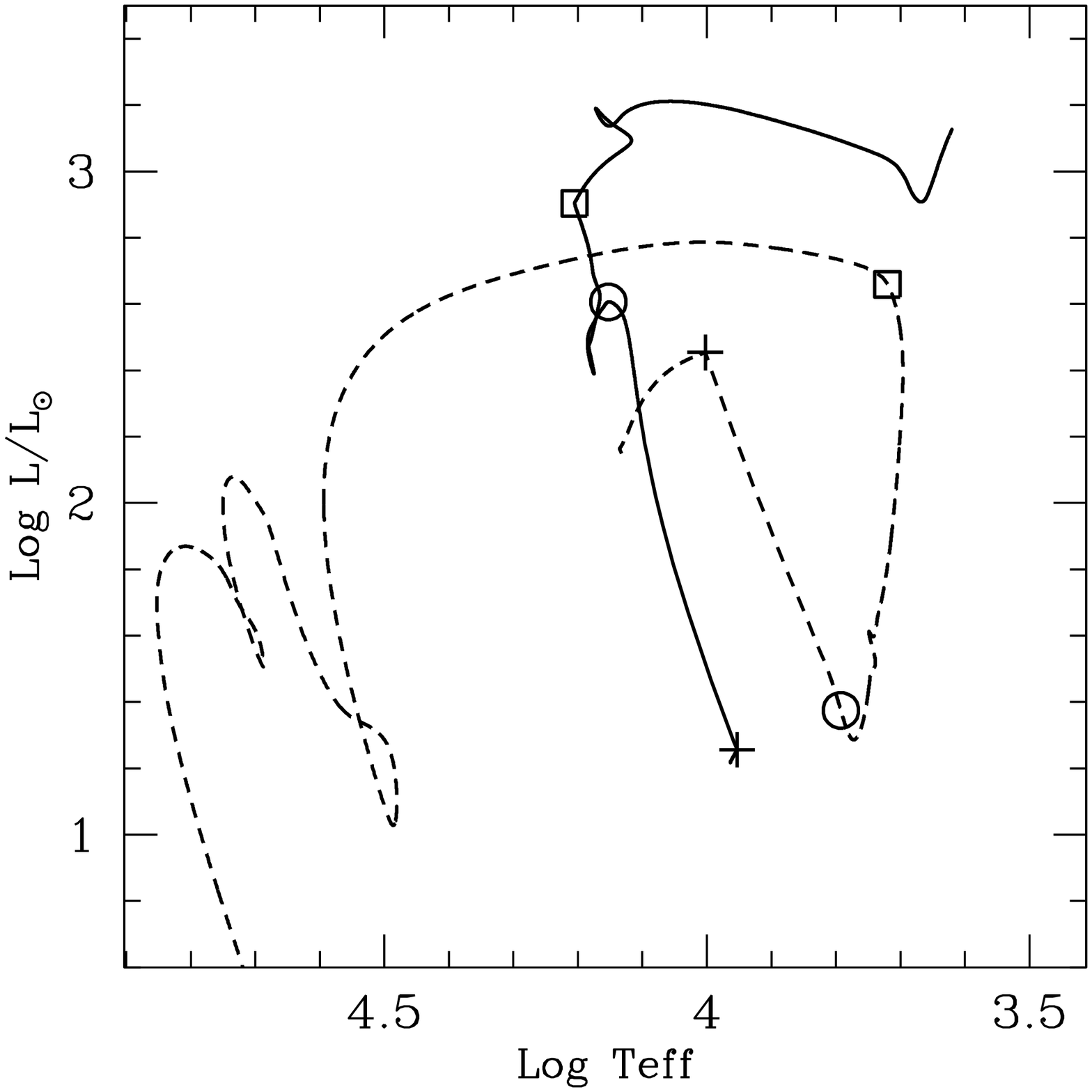}
\hfill
\includegraphics[width=45mm,height=75mm]{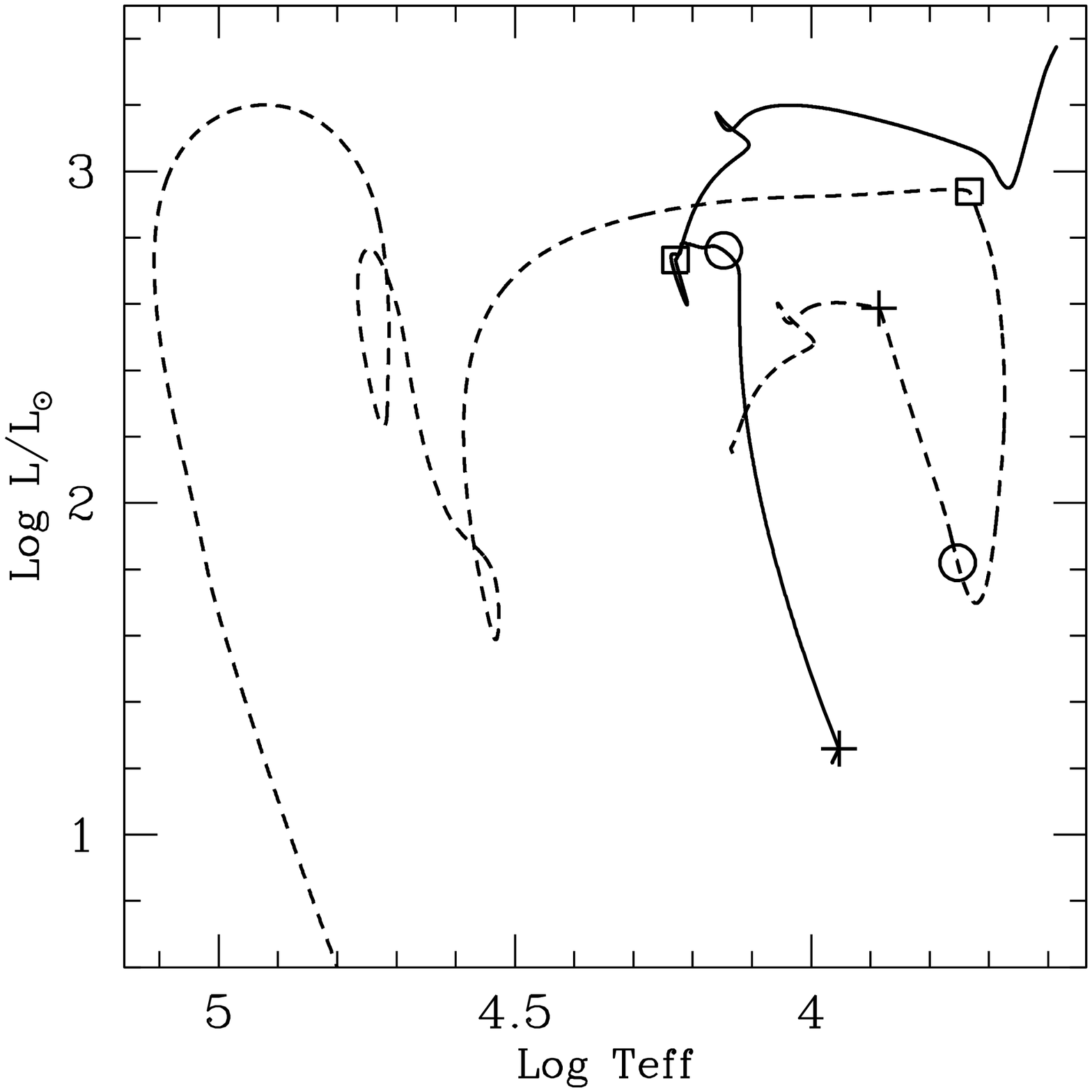}
\hfill
\includegraphics[width=45mm,height=75mm]{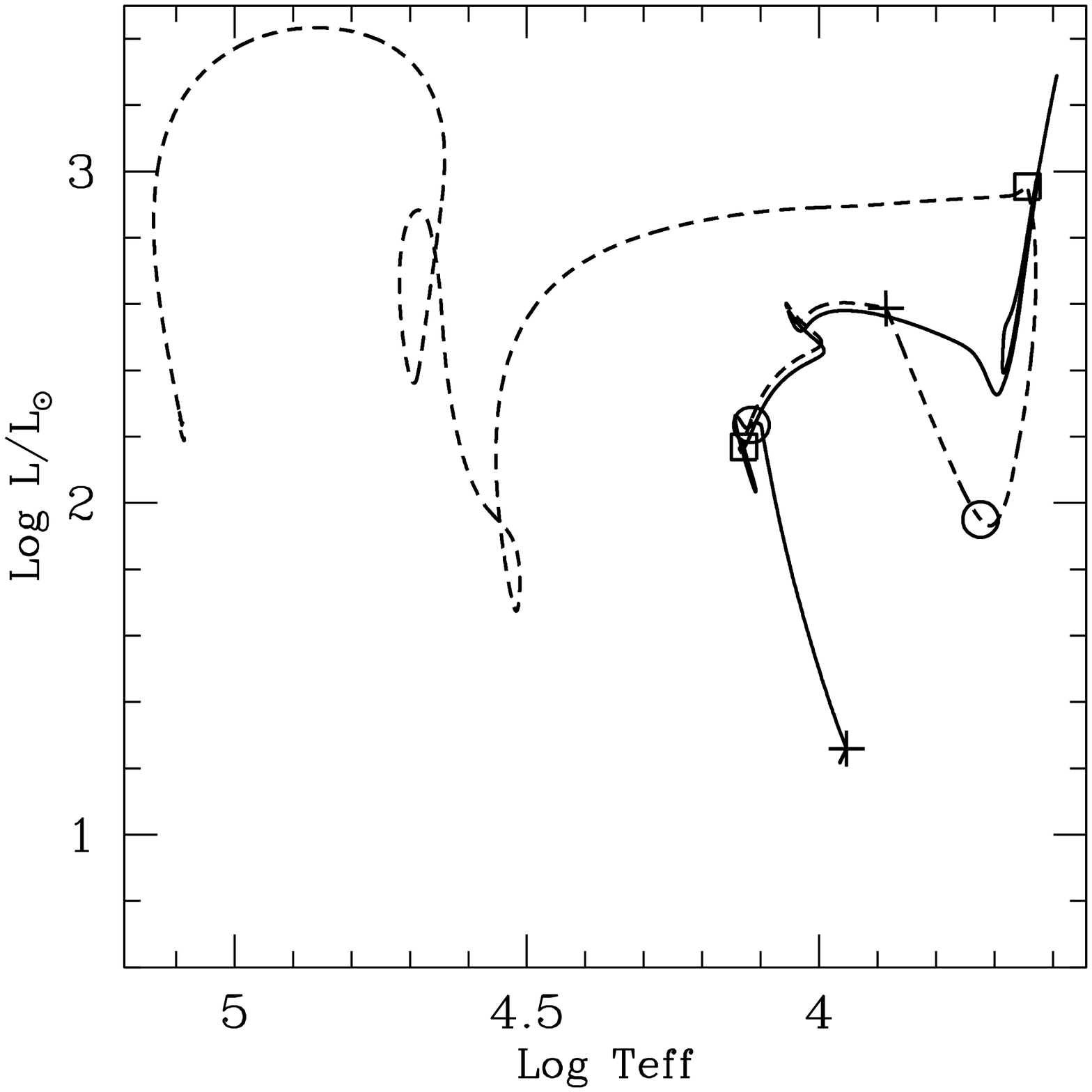}

\caption{The evolutionary tracks of two components (3.5 + 2.0
$M_\odot$) undergoing case A(the left panel) and case B scheme
($\beta$=100\% for the middle panel and $\beta$=50\% for the right
panel). The dashed and solid lines are the evolutionary tracks of
the primary and the secondary respectively. The plus signs, open
circles, open squares show their positions, in order, when: the mass
transfer begins, the mass ratio equals 1, the mass transfer
terminates. \label{f:35hr}}
\end{figure}

We use the stellar evolutionary code of Eggleton (1971,1972,1973)
which is updated by Han et al. (1994, 2000) \& Pols et al. (1995,
1998) to compute close primordial binary evolutionary models instead
of using analytic formulae to approximate the main characters of
stellar evolutions. Some inner stellar parameters are needed to be
taken care because they have tight relations to the unsolved
problems in the theory of stellar structure and evolution,
especially for the massive stars (Deng et al. 2001; Tian et al.
2009). The Eggleton's code can provide more detailed information on
the evolution of close binaries. The code uses the radiative opacity
library from Iglesias \& Rogers (1996) and molecular opacities of
Alexander \& Ferguson (1994). Roche lobe overflow (RLOF) is treated
as a boundary condition within the code. The accreted material from
the primary is assumed to be simply deposited on to the surface of
the secondary and distributed homogeneously all over the outer
layers instantly.

We consider two main mechanisms of BS formation in this work which
mainly depend on the initial mass ratio. During RLOF, the secondary
may accrete material transferred from the primary or merge with the
primary depending on the initial mass ratio q($M_2$/$M_1$). With a
small $q<q_{\rm crit}$, RLOF is dynamical unstable and the secondary
quickly fills its Roche lobe at the onset of mass transfer. A common
envelope (CE) is formed afterward. The CE may be ejected depending
on its binding energy or the binary will merge. The remanet could be
a BS if both companions are main sequence stars. The merge process
is complicated and the physic during the process is still uncertain.
The composition of the mergers may influence their lifetimes on MS
and their positions on CMD whereas the fully mixed model is an
extreme case. Any alternative will lead to a shorter MS lifetime. We
constructed the merger models by assuming all the low q mergers are
fully mixed and possess the corresponding binary mass. With a larger
$q>q_{\rm crit}$, RLOF is quite stable and can be calculated by
Eggleton's code. If the secondary is a main-sequence star, the
luminosity will goes upward along the main sequence as a result of
mass accretion. Whether it could become a BS or not depends on the
final mass when RLOF terminates. For some case A models, the system
may become contact during RLOF because of the small initial space
orbital separation. Previous studies indicate that contact binary
will eventually coalesce (Webbink 1976; Eggleton 2000; Li, Han \&
Zhang 2005) although there are still some debates over the
coalescence time-scale (Eggen \& Iben 1989; van't Veer 1994;
Dryomova \& Svechnikov 2002; Bilir et al. 2005). The remnants of
binary coalescence could contribute to luminous BSs, however their
number is very limited to make a notable influence to the
statistical property of clusters (Chen \& Han 2009). The number of
this case in our model is also very limited so we just terminate the
code when the two components come into contact. We are unable to
follow the RLOF that begins during red giant branch (RGB) owning to some
numerical problems. Finally, we calculate a series of close binary
evolutionary models covering case A and part of case B (RLOF during
Hertzsprung gap (HG)) with the initial primary mass $M_1$ = 2.0 - 7.6
$M_\odot$ with solar metallicity. The interval of $M_1$ is 0.1 $M_\odot$
when $M_1 < 6.0 M_\odot$ and 0.2 $M_\odot$ when $6.0 M_\odot < M_1 < 7.6 M_\odot$.
The initial mass ratios $M_2/M_1$ are from 0.2 to 0.9 with the increment of 0.1.
The initial orbital separations are from 5.0 $R_\odot$ to 320.0 $R_\odot$
with the increment of 2.0 $R_\odot$ for most of our models
and 4.0 $R_\odot$ only when $M_1 > 5.0 M_\odot$ and
$R_{\rm initial} > 40.0 R_\odot$. More details about the modeling of
the evolutionary models and the configurations of the code have been
discussed in LDZ10.

Fig. 1 shows the evolutionary tracks of two close binary systems
(3.5$M_\odot$+2.0$M_\odot$). On the left panel the initial orbital
is 13.0 $R_\odot$ therefor follows case A mass transfer while the
other two have the orbital separations doubled and go through case B
mass transfer (the middle and the right panel). Different mass
transfer rates are set in both case B plots and this issue will be
discussed in the next paragraph. Solar composition (Z = 0.020, Y =
0.280) is adopted in the models . The mass transfer begins when the
primary overfills its Roche lobe at MS for case A (the left panel)
and HG for case B (the middle and the right panel), and slows down
as the primaries evolve to the bottom of the RGB in both models. The
mass ratio quickly reverses when reaching the bottom of the RGB. As
a result of mass accretion, the luminosity of the secondaries evolve
upward along the main sequence in CMD as shown by the solid lines on
all the plots. After the termination of mass transfer, the accretors
in the models follow the regular evolutionary paths of single stars
with about 90 percent of the total system mass. The previous mass
donors become C-O dwarfs with $\sim0.44$ $M_\odot$ in the first case
(case A mass transfer) and $\sim0.58$ $M_\odot$ in the second one
(case B mass transfer).

Conservations in both mass and angular momentum are assumed for case
A and case B. However the assumption is only reasonable for a
restricted range of intermediate-mass binaries as predicted by
Nelson \& Eggleton (2001). The efficiency of mass transfer, $\beta$,
which defined as the mass fraction of the matter accreted by the
secondary to that lost from the primary, is one of the major
uncertainties in the evolutionary calculations of close binaries. In
general, the factor of $\beta$ could influence the final mass of the
secondary after RLOF. Many relevant processes are still not yet well
understood and the value is therefore unclear. Obviously, with a
larger $\beta$, the companion could accrete more material and become
more massive and therefor more luminous. However, the lifetime of
the secondary on main sequence after RLOF decreases along with the
increasing final mass. The larger $\beta$ the shorter it remains on
MS. Usually a high value is suggested for the mass donor on MS or HG
and low value for the mass donor on first giant branch(FGB) or
asymptotic giant branch(AGB). De Mink et al.(2007) presented 20000
detailed evolution models to fit a sample of 50 double-lined
eclipsing binaries in the Small Magellanic Cloud. They found no
single value of $\beta$ that can explain all the systems. In many
calculations of binary evolution, usually a single constant mass
transfer efficiency of 0.5 has been assumed (e.g. De Greve \& De
Loore 1992, Chen \& Han 2002,2009). In order to investigate the
non-conservative case which is more likely in reality, we consider
such an effect by using a constant mass transfer of efficiency. The
right panel in Fig. 1 is an example. Due to the mass loss from the
system during RLOF, the luminosity of the accretor can not reach as
high as the conservative case on HR diagram. A separate set of
non-conservative models are calculated using the same parameter set,
the results of which will be discussed in Sect 3.

\begin{figure}
\includegraphics[scale=0.35]{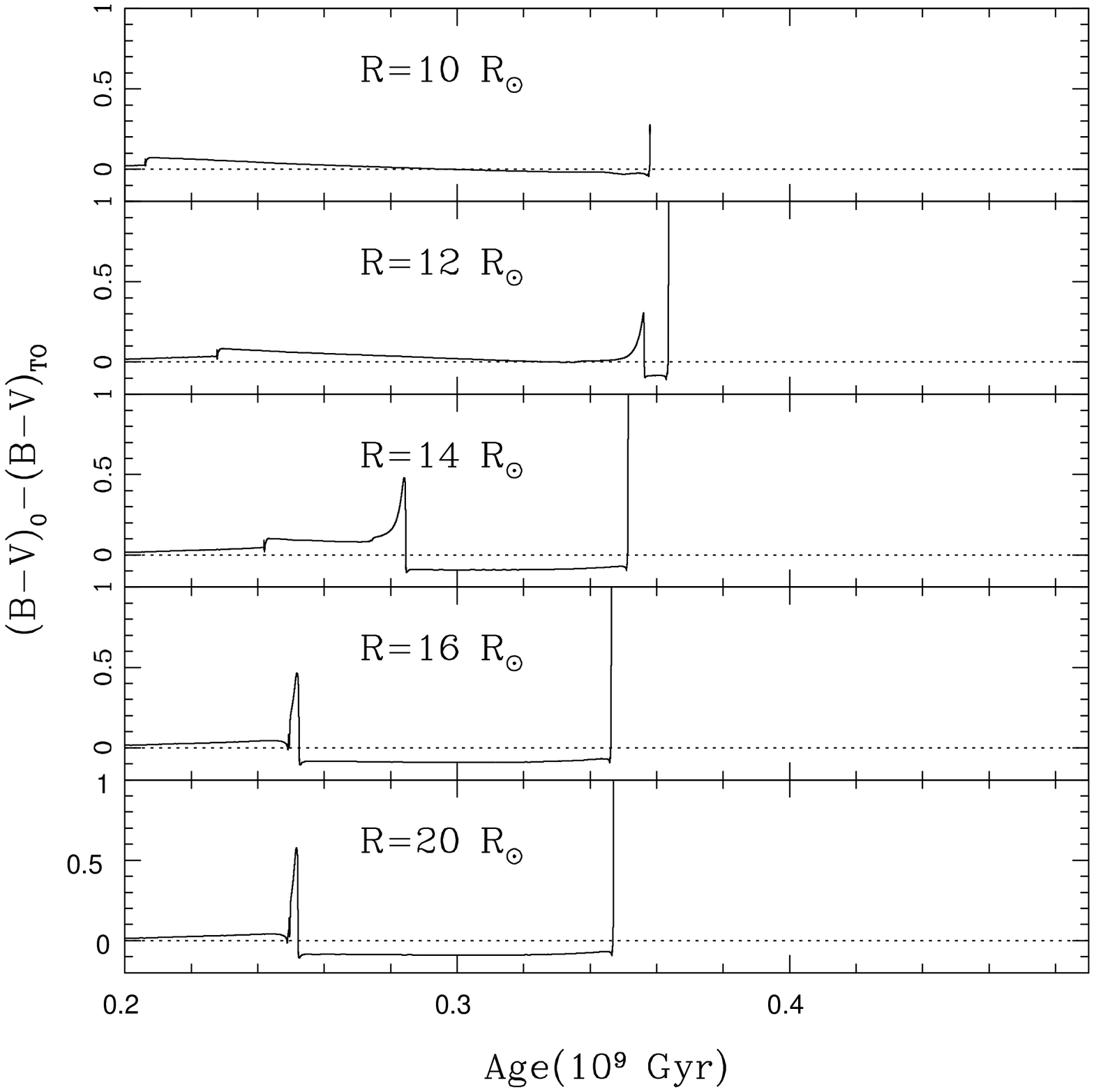}
\hfill
\includegraphics[scale=0.35]{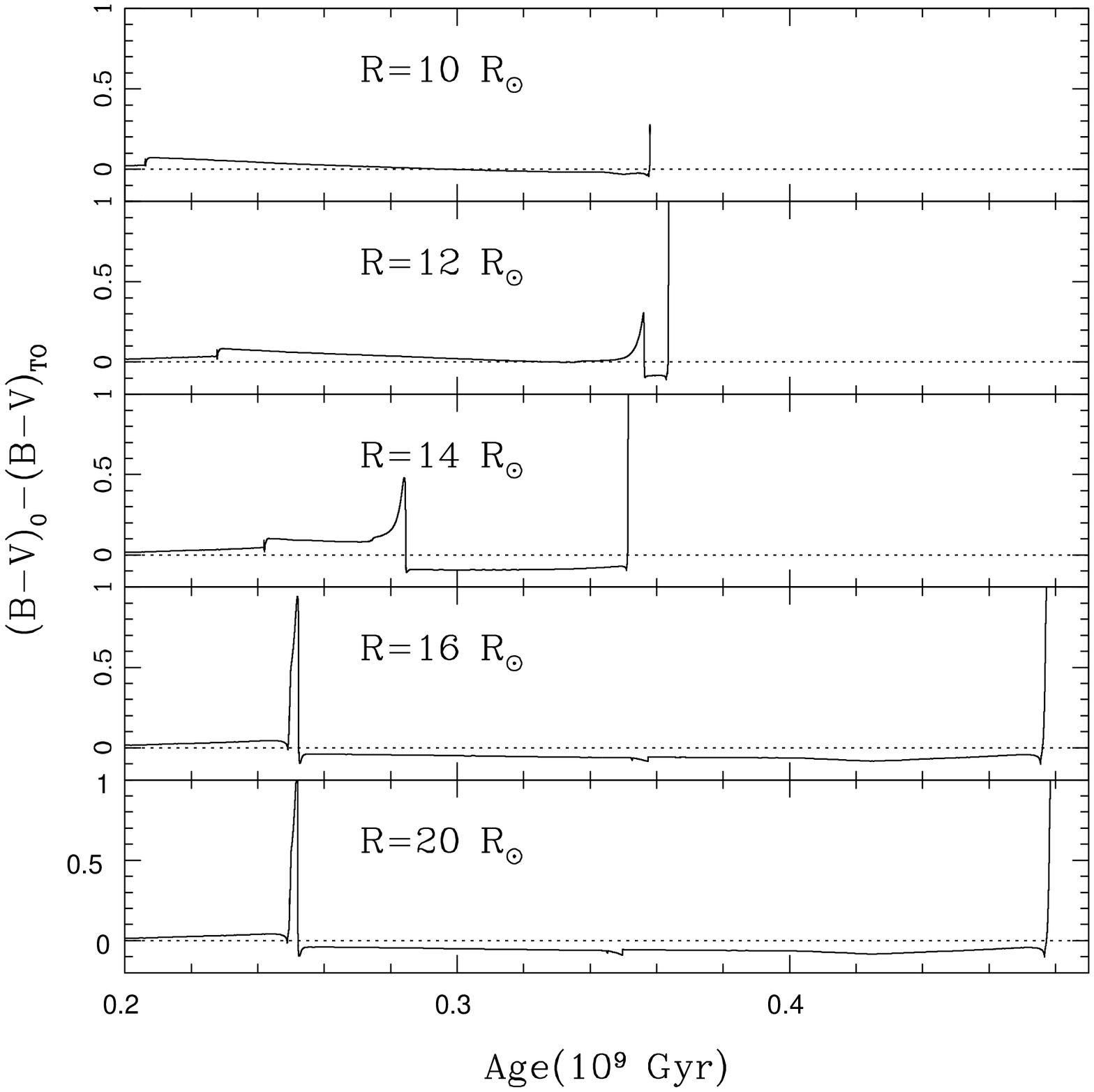}
\caption{Evolution of synthetic color for example models with $M_1=
3.5M_\odot$, $M_2=2.0M_\odot$ at 5 different orbital separations.
Case B models on the left panel are $\beta$=1 while those on the
right panel are $\beta$=0.5. The dotted lines indicate
$(B-V)_0-(B-V)_{\rm TO}=0$.\label{f:35color}}
\end{figure}

The time of the on-set of mass transfer is critical for the
subsequent evolution as shown in our previous study (LDZ10). Our
study shows that the duration and luminosity of close binary systems
during and after mass transfer vary greatly. Fig 2. shows the
synthetic colors of the system with respect to that of the turn-off
of the corresponding isochrones at the same age as a function of
age. The initial parameters of the systems in the figure are:
$M_1$=3.5$M_\odot$, $M_2$=2.0$M_\odot$ and 5 different initial
orbital separations of 10, 12, 14, 16 and 20 $R_\odot$. The
conservative cases are shown on the left panel and the
non-conservative cases are shown on the right panel. The upper 3
rows in both plots are case A models and the bottom 2 are case B
models. The dotted lines ($(B-V)_0-(B-V)_{\rm TO}=0$) is the location of
the TO point at different ages. The portions of the solid lines
below the dotted lines are BS phase. We can notice clearly that the
durations of binary models being in BS phase region become longer
while the initial orbital separation increases; and case B makes the
evolution of the binary systems much longer in BS phase than case A
and even longer in the non-conservative case as shown at the bottom
2 rows on the right panel. The evolutionary behaviors of case B
models are similar when $R_{\rm initial}>16R_\odot$. We can also notice
that BSs formed via case B seem brighter and bluer than those formed
via case A. This behavior is similar to binary systems with lower
masses as in LDZ10.

\section{Monte-Carlo simulations}
\label{sect:simu}

It is possible to investigate the statistical property of BS
population from mass transfer and merger based on our models. We
have tested the detailed method in a previous work to simulate the
BS population in old clusters such as M67. In this work, we would
like to extend the algorithm to cover younger clusters and make it
possible to compare our theoretical models of BS populations to the
observations (Ahumada et al. 2007, hereafter AL07). We performed a
series of Monte-Carlo simulations for a large sample of $10^6$
binaries. A single starburst is assumed in our calculation. The
initial mass function (IMF) of the primary, the initial mass ratio
distribution and the distribution of initial orbital separations are
the same as in our previous studies for clarity. They are also
described as follows.

1.The IMF of Kroupa et al. (1991) which reproduces a distribution of
masses similar to the best fits of KTG is adopted. It agrees well in
U-B and B-V color with the traditional way of constructing IMF of
KTG93 on SSPs (Zhang et al. 2005). It is convenient to use in
Monte-Carlo simulation and the initial mass of the binaries can be
given by the generating function.

\begin{equation}
  M(X) = 0.33[\frac{1}{(1-X)^{0.75}+0.04(1-X)^{0.25}}-\frac{(1-X)^2}{1.04}]
  \label{phipsi}
\end{equation}
where $X$ is a random number uniformly distributed between 0 and 1.
$M(X)$ is the binary mass in units of $M_\odot$. $M(X)$ is limited
between 0.2 to 100.0 $M_\odot$ by the assumption of single-star
population covering 0.1 to 50.0 $M_\odot$.

2.We adopt an uniform distribution of mass ratio as given by Hurley
et al. (2001)
\begin{equation}
  1 > q > \rm Max[\frac{0.1}{M(X)-0.1},0.02(M(X)-50.0)]
  \label{phipsi}
\end{equation}

3.We assume the distribution of separations is constant in log
$\alpha$ ($\alpha$ is the orbital separation) from
Pols \& Marinus (1994). The lower limit of separation is
where a ZAMS star fills its Roche lobe and the
upper limit adopted here is 50 au (Hurly et al. 2005).

\section{Results and Discussions}

The current $M_{\rm V}$ and $(B-V)_0$ of any close binaries whose
initial parameters are in the range of our grid can be obtained by
linear interpolations. The simulation space has 3 degrees of
freedom: orbital separation, mass ratio and donor mass. The
parameters are defined in a huge table(available online). Table
1\footnote{The complete table and the evolutionary tracks of all the
models are only available in electronic form via
http://sss.bao.ac.cn/BS-model2/} is a truncated version for
demonstration.

\begin{table}
 \centering
 \begin{minipage}{80mm}
\caption{Models of case A, case B and mergers. Only a sample of the
table is included here. The full table is available in electronic
form. }\label{t:model}
\begin{tabular}{@{}ccccccc@{}}
  \hline
   $\rm Log(age)$ & $M_1$   &   $M_2$   &     $\alpha$     & $V$ & $B-V$ & $\beta$ \\
 yr & ($M_\odot$) & ($M_\odot$) & ($R_\odot$) &     &     & \\
  \hline
 8.25 & 4.5 & 2.7 & 10.0  &  -2.241  & -0.116 &  1.0 \\
 8.25 & 4.5 & 2.7 & 12.0  &  -2.161  & -0.108 &  1.0 \\
 8.25 & 4.5 & 2.7 & 14.0  &  -2.113  & -0.194 &  1.0 \\
 8.25 & 4.5 & 2.7 & 16.0  &  -2.349  & -0.185 &  1.0 \\
 8.25 & 4.5 & 2.7 & 18.0  &  -2.431  & -0.180 &  1.0 \\
 8.25 & 4.5 & 2.7 & 20.0  &  -2.425  & -0.180 &  1.0 \\
 8.25 & 4.5 & 2.7 & 22.0  &  -2.415  & -0.180 &  1.0 \\
 8.25 & 4.5 & 2.7 & 24.0  &  -2.407  & -0.180 &  1.0 \\
 8.25 & 4.5 & 2.7 & 26.0  &  -2.400  & -0.180 &  1.0 \\
 8.25 & 4.5 & 2.7 & 28.0  &  -2.394  & -0.181 &  1.0 \\
 8.25 & 4.5 & 2.7 & 30.0  &  -2.387  & -0.181 &  1.0 \\
 8.25 & 4.5 & 2.7 & 32.0  &  -2.382  & -0.181 &  1.0 \\
 8.25 & 4.5 & 2.7 & 34.0  &  -2.377  & -0.181 &  1.0 \\
 8.25 & 4.5 & 2.7 & 36.0  &  -2.374  & -0.181 &  1.0 \\
 8.25 & 4.5 & 2.7 & 38.0  &  -2.369  & -0.181 &  1.0 \\
 8.25 & 4.5 & 2.7 & 40.0  &  -2.366  & -0.181 &  1.0 \\
 8.25 & 4.5 & 2.7 & 42.0  &  -2.364  & -0.181 &  1.0 \\
\hline
\end{tabular}
\end{minipage}
\end{table}

\begin{table*}
 \centering
 \begin{minipage}{140mm}
\caption{Parameters of young clusters with metallicity similar to
the Sun (Z = 0.02)}\label{t:Parameter}
\begin{tabular}{@{}cccccccccc@{}}
  \hline
   $\rm Cluster Name$   &   $\rm log(t/yr)$   &     $E(B-V)$     & $DM$ & $Z$ & $[Fe/H]$ & $N_{2}$ & $N_{\rm BS}$ & $f$ & $\rm References$ \\
 (1) & (2) & (3) & (4) & (5) & (6) & (7) & (8) & (9) \\
  \hline
 NGC 2516    & 8.15 & 0.12 &  7.93 & 0.018 & -0.05 &   40(35) & 1(6) & 0.025(0.171) & 1,2 \\
 NGC 5316    & 8.20 & 0.27 & 11.26 & 0.019 & -0.02 &   10(20) & 0(4) & 0.000(0.200) & 1,4,6 \\
 IC 2488     & 8.25 & 0.24 & 11.20 & 0.019 & -0.02 &   30(10) & 3(1) & 0.100(0.100) & 1,17 \\
 NGC 1545    & 8.45 & 0.30 & 10.19 & 0.017 & -0.06 &   10(20) & 0(1) & 0.000(0.050) & 1,4,6 \\
 NGC 6281    & 8.50 & 0.15 &  8.93 & 0.02  &  0.00 &   20(25) & 0(4) & 0.000(0.160) & 1,4,6 \\
 IC 2714     & 8.50 & 0.36 & 11.68 & 0.015 & -0.12 &  110(80) & 2(1) & 0.018(0.013) & 1,14 \\
 NGC 1027    & 8.55 & 0.33 & 10.46 & 0.023 &  0.06 &   40(40) & 0(2) & 0.000(0.050) & 1,4,5 \\
 Melotte 111 & 8.60 & 0.00 &  4.77 & 0.019 & -0.03 &   10(10) & 1(1) & 0.100(0.100) & 1,8 \\
 NGC 1245    & 8.70 & 0.30 & 12.27 & 0.018 & -0.05 &  160(75) & 7(9) & 0.044(0.120) & 1,9,10 \\
 NGC 6633    & 8.80 & 0.17 &  7.80 & 0.015 & -0.11 &   40(40) & 4(3) & 0.100(0.075) & 1,11 \\
 NGC 2539    & 8.80 & 0.06 & 10.60 & 0.018 & -0.04 &  100(20) & 1(1) & 0.010(0.050) & 1,19,20 \\
 NGC 2477    & 8.85 & 0.28 & 10.50 & 0.018 & -0.05 & 330(190) &15(28)& 0.045(0.147) & 1,15 \\
 IC 4756     & 8.90 & 0.23 &  7.60 & 0.022 &  0.04 &   80(55) & 6(1) & 0.075(0.018) & 1,11 \\
 NGC 5823    & 8.90 & 0.50 & 10.50 & 0.016 & -0.10 &   35(35) & 1(1) & 0.029(0.029) & 1,4,16 \\
 Collinder 223& 8.00& 0.25 & 13.00 & 0.02? &   ... &   25(25) & 2(2) & 0.080(0.080) & 1,21 \\
 NGC 5617    & 8.15 & 0.54 & 11.53 & 0.02? &   ... &   65(70) & 0(9) & 0.000(0.129) & 1,3 \\
 NGC 5460    & 8.30 & 0.144&  9.49 & 0.02? &   ... &   20(20) & 0(1) & 0.000(0.050) & 1,12 \\
 Melotte 105 & 8.40 & 0.52 & 11.80 & 0.02? &   ... &   80(25) & 4(1) & 0.050(0.040) & 1,7 \\
 NGC 2383    & 8.60 & 0.22 & 13.30 & 0.02? &   ... &    5(5)  & 1(1) & 0.200(0.200) & 1,13 \\
 NGC 2818    & 8.70 & 0.22 & 12.90 & 0.02? &   ... &   45(45) & 2(5) & 0.044(0.111) & 1,18 \\
\hline
\end{tabular}
\end{minipage}
\begin{minipage}{160mm}
Notes. Columns (1)-(6) are cluster name, age, color
excess, distance modulus, metallicity (Z) and [Fe/H]. Columns
(7)-(9) are $N_2$ (number of stars within 2 mag below the
main-sequence turn-off stars), $N_{\rm BS}$ in the sample clusters from
observations and the frequency $f$=$N_{\rm BS}/N_2$(the data in brackets
are from Xin07). The last column is the references for the data.
(1)AL07; (2)Terndrup et al.2002; (3)Ahumada 2005; (4)Kharchenko et
al.2005; (5)Loktin \& Matkin 1994; (6)Dias et al. 2002; (7)Sagar et
al. 2001; (8)van Leeuwen 1999; (9)Burke et al. 2004; (10)Subramaniam
2003; (11)Hebb et al. 2004; (12)Barrado \& Byrne 1995;
(13)Subramaniam \& Sagar 1999; (14)Claria et al.1994; (15)Eigenbrod
et al.2004; (16)Janes 1981; (17)Claria et al.2003; (18)Surendiranath
et al. 1990; (19)Marshall et al.2005; (20)Lapasset et al.2000
(21)Tadross 2004.
\end{minipage}
\end{table*}

\begin{figure}
\includegraphics[scale=0.40]{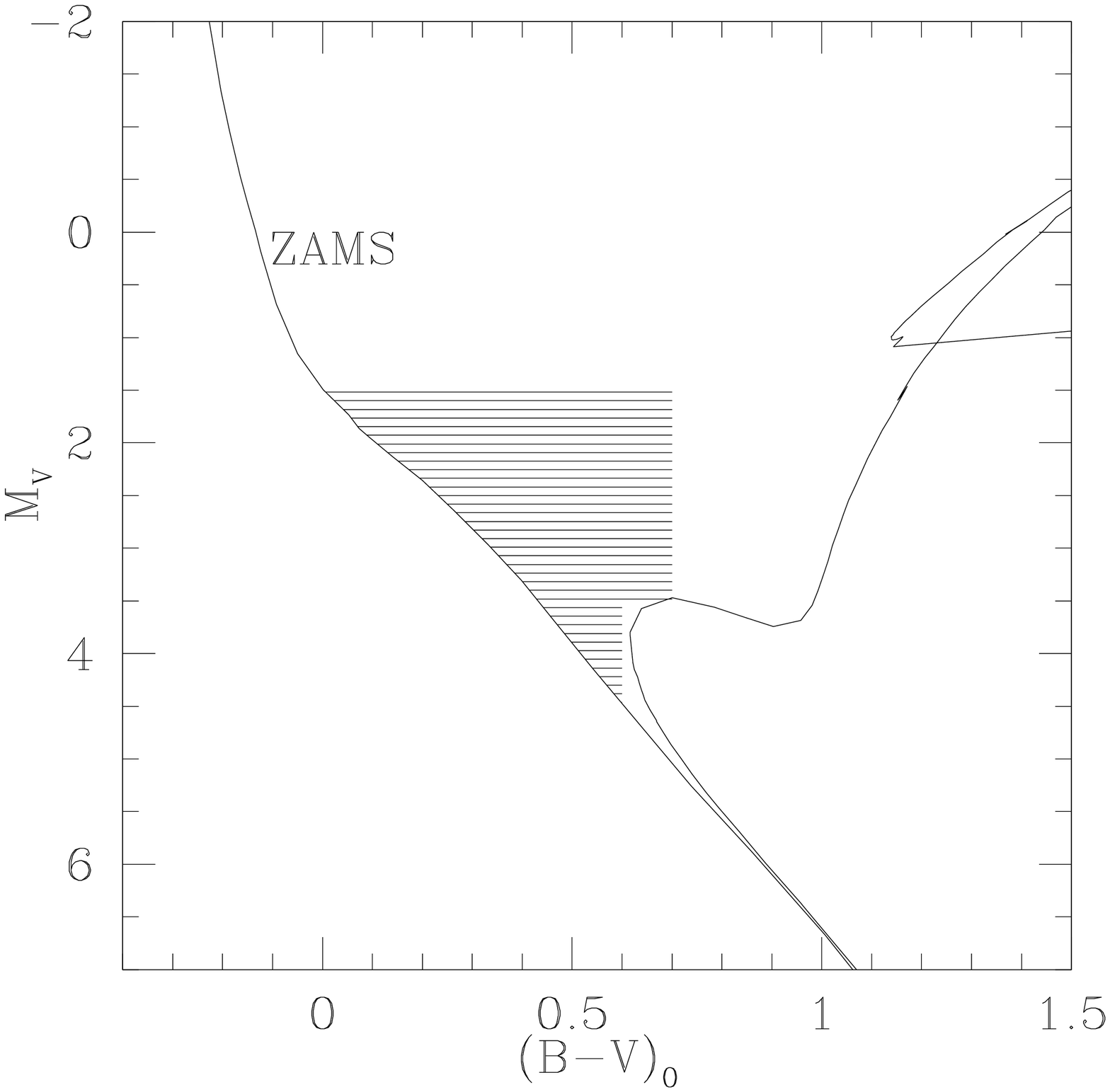}
\hfill
\includegraphics[scale=0.40]{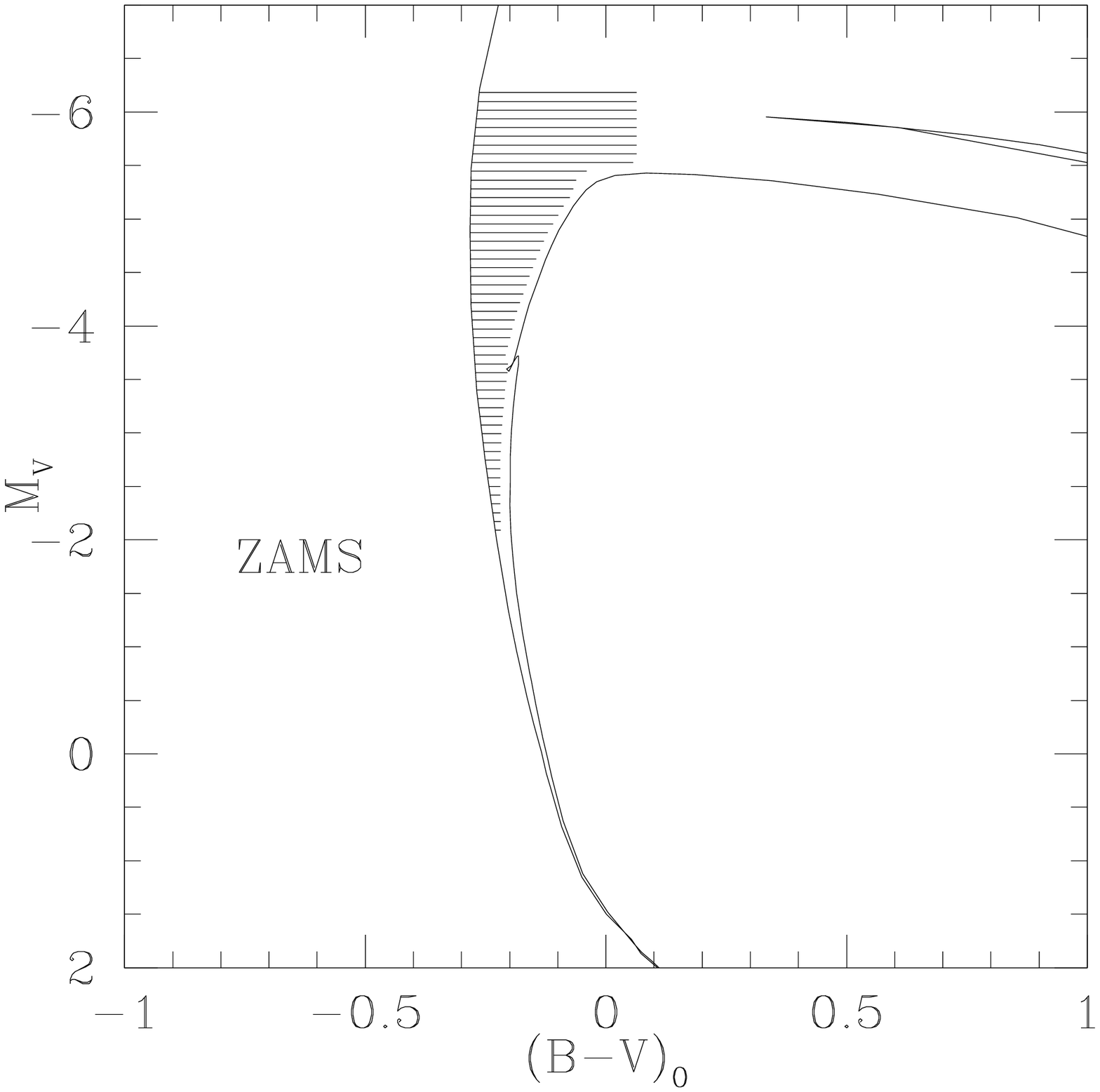}
\caption{CMDs for an old open cluster and a young open cluster: the
isochrones correspond to log(age) = 9.8 for the left and log(age) =
7.5 for the right. The blue straggler areas are shaded.
\label{f:def}}
\end{figure}

\begin{figure*}
\includegraphics[scale=0.24]{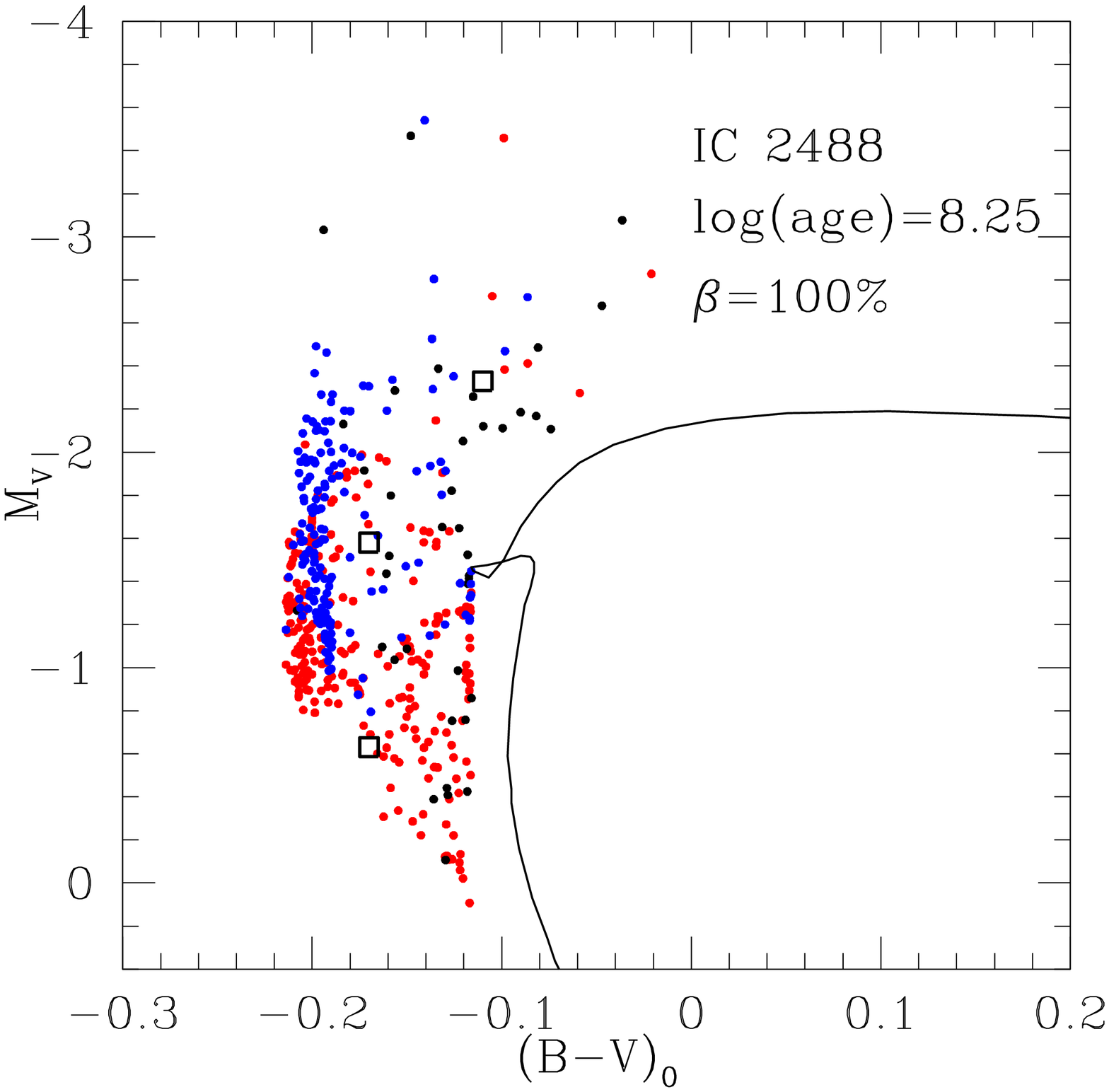}
\hfill
\includegraphics[scale=0.24]{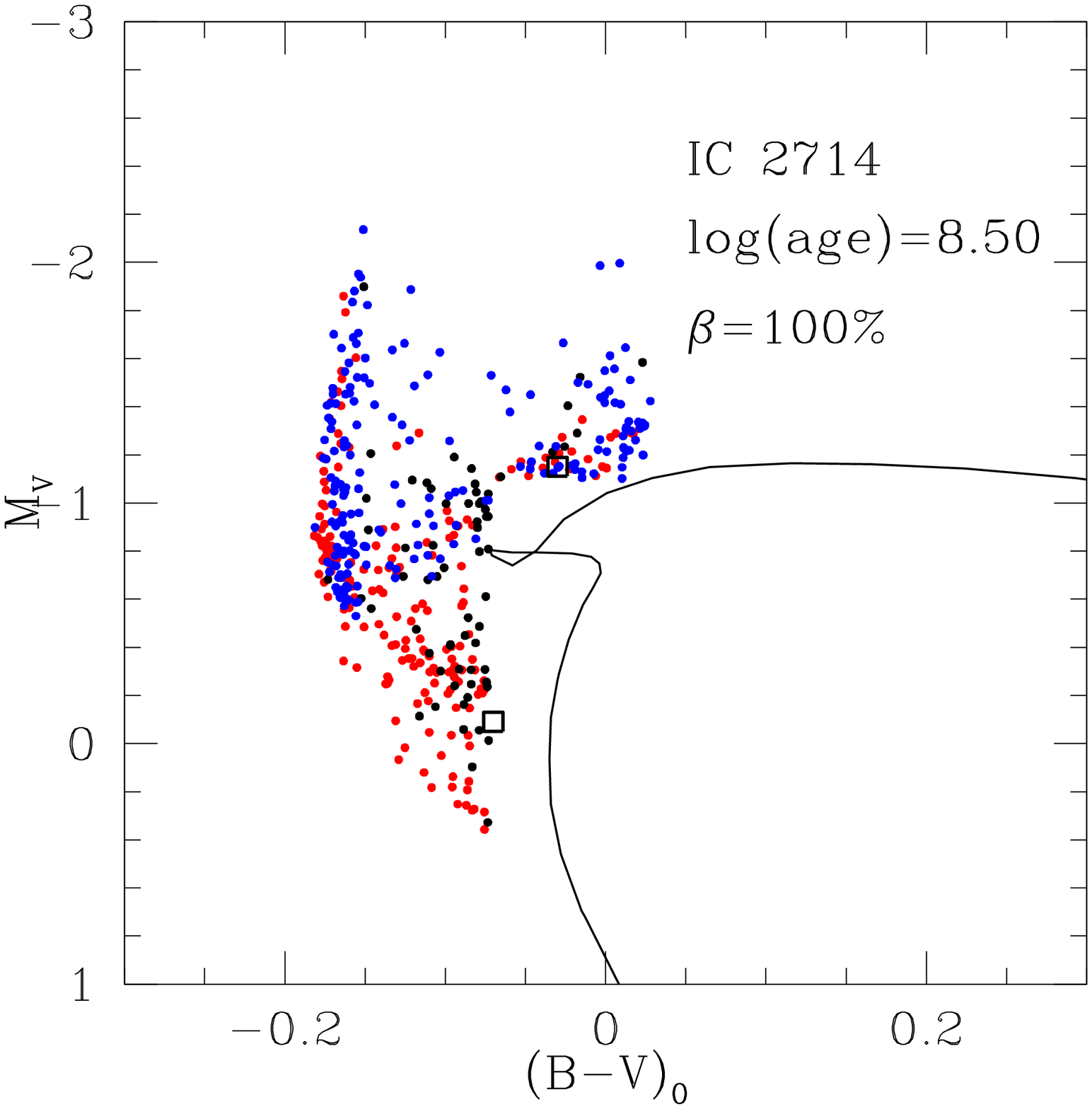}
\hfill
\includegraphics[scale=0.24]{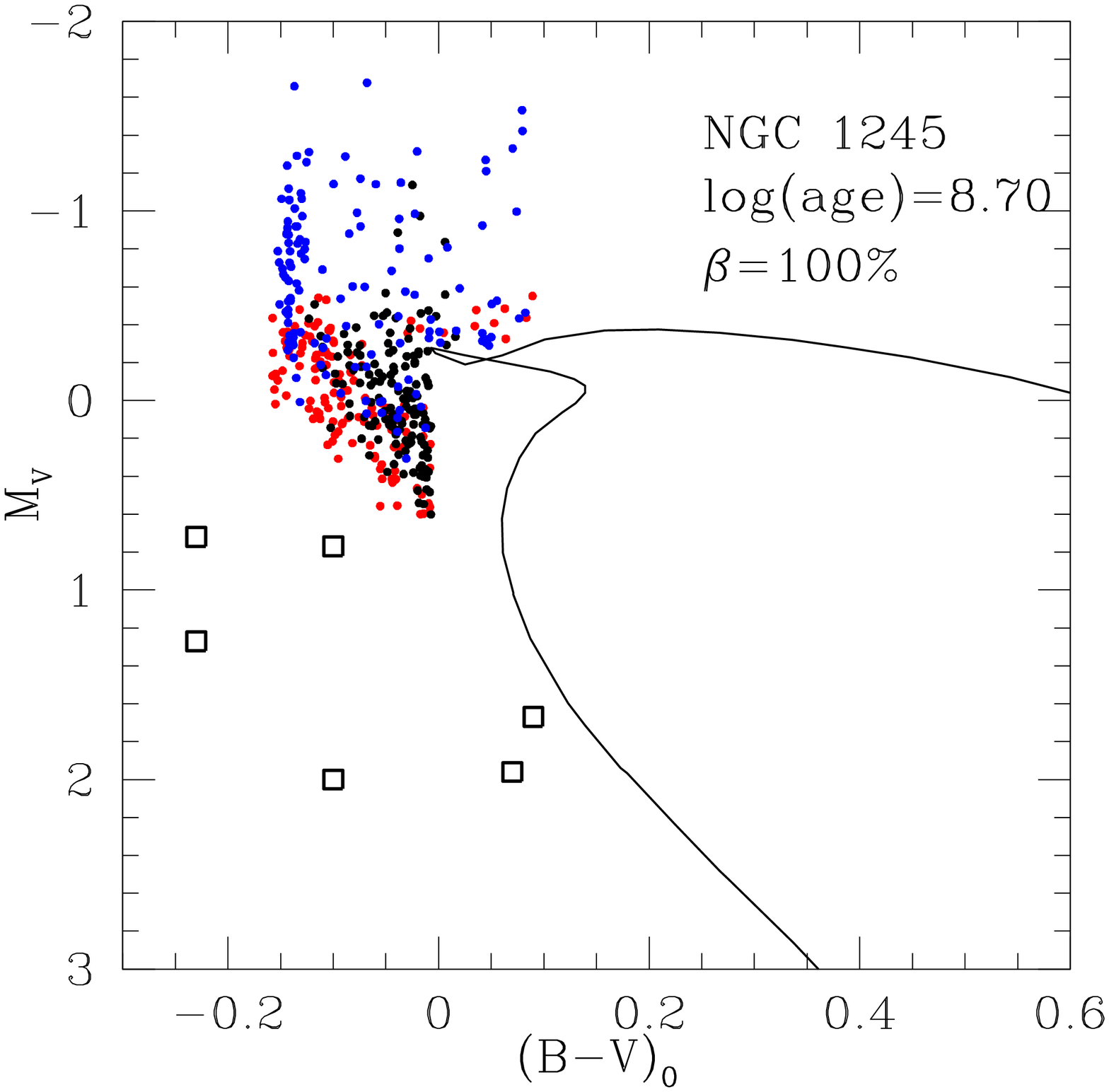}
\hfill
\includegraphics[scale=0.24]{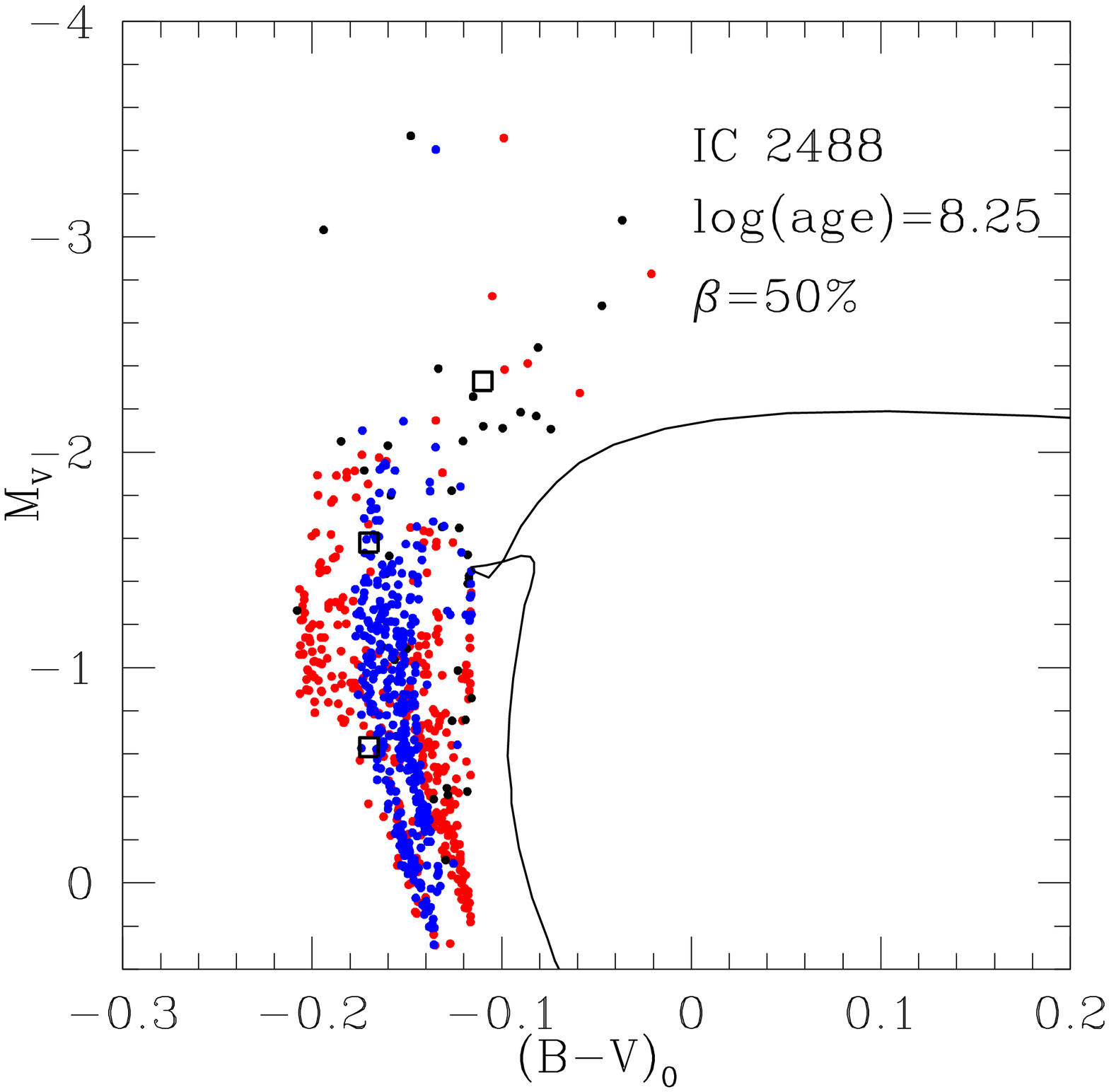}
\hfill
\includegraphics[scale=0.24]{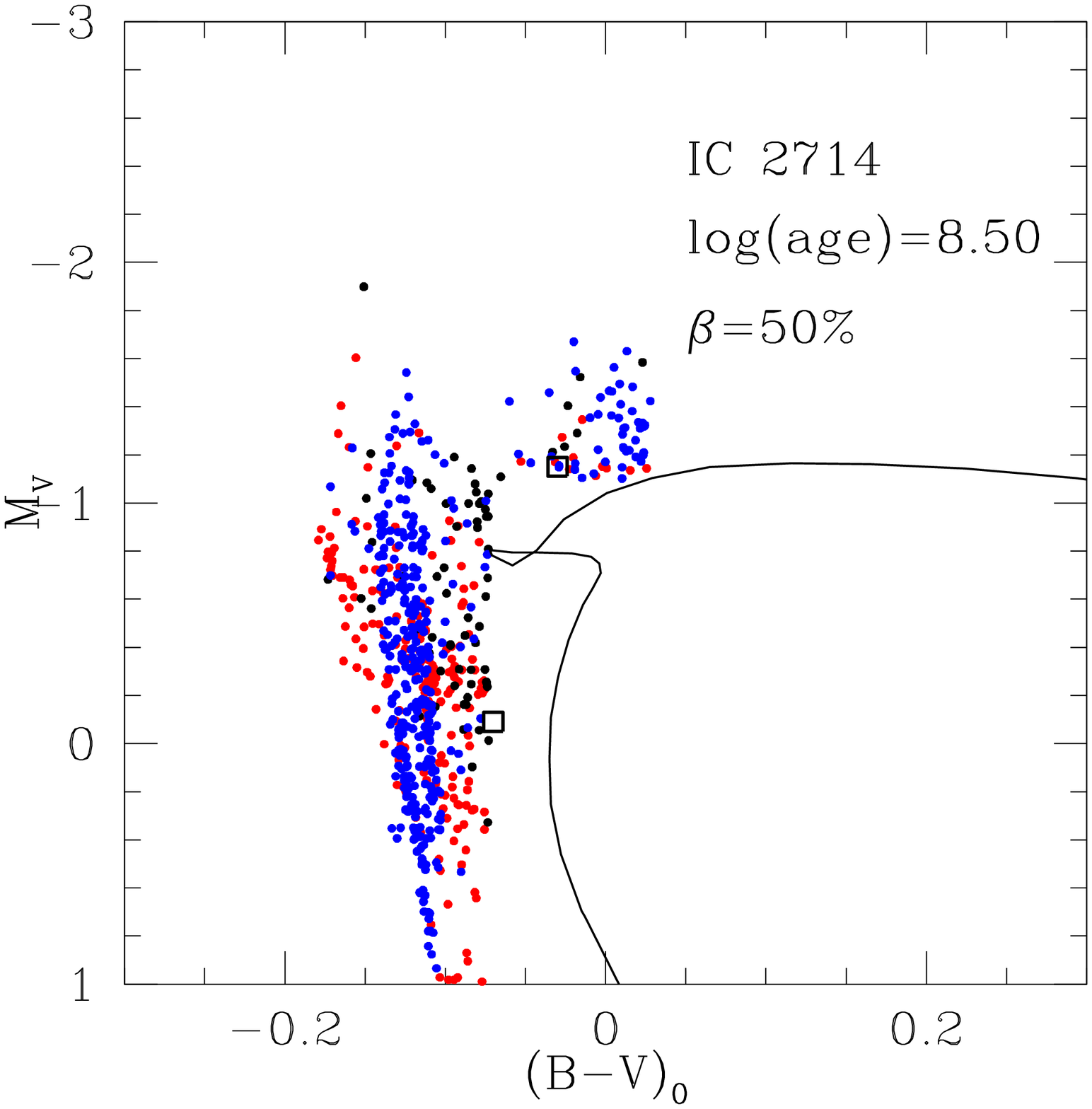}
\hfill
\includegraphics[scale=0.24]{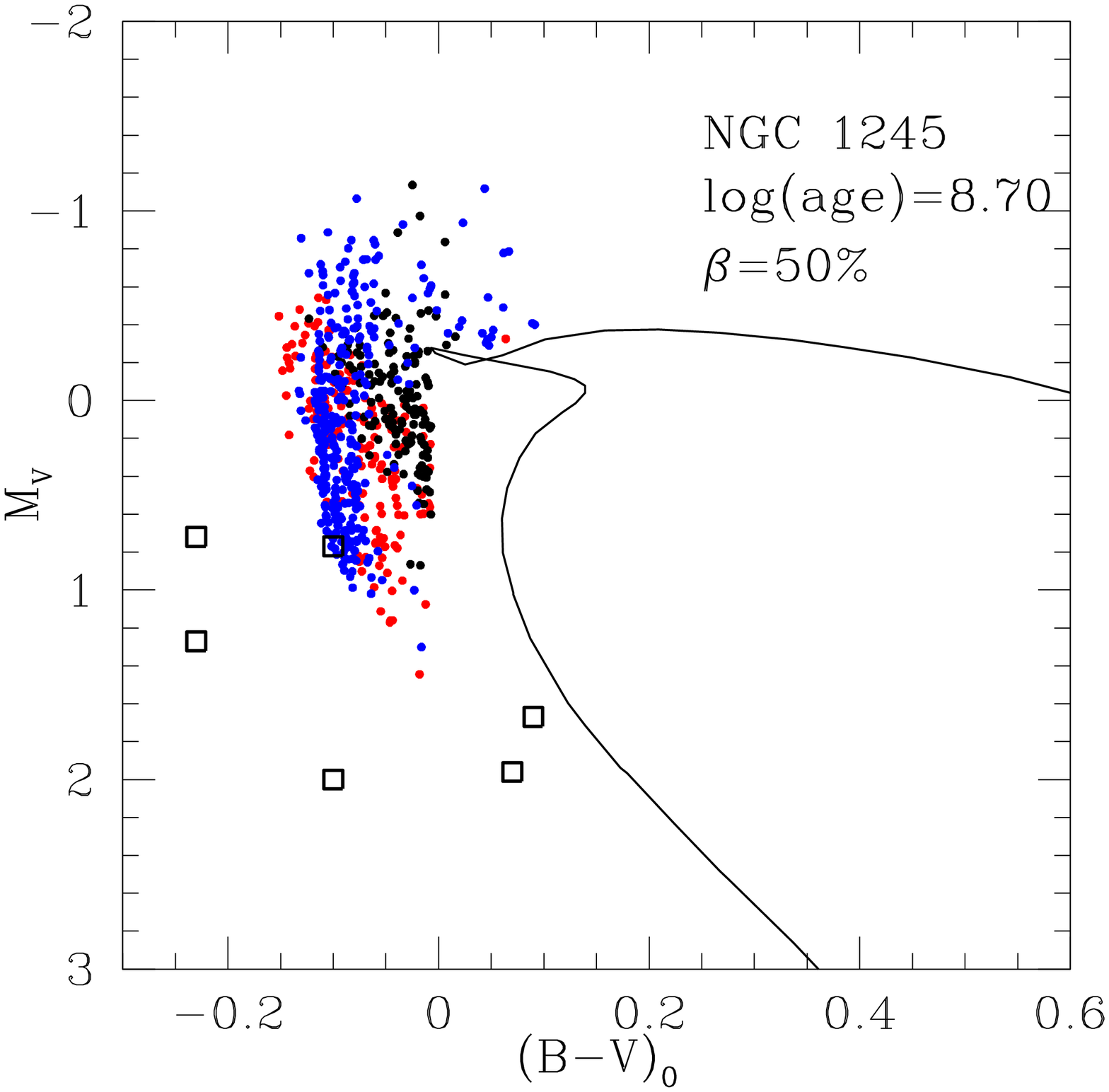}
\hfill
\includegraphics[scale=0.24]{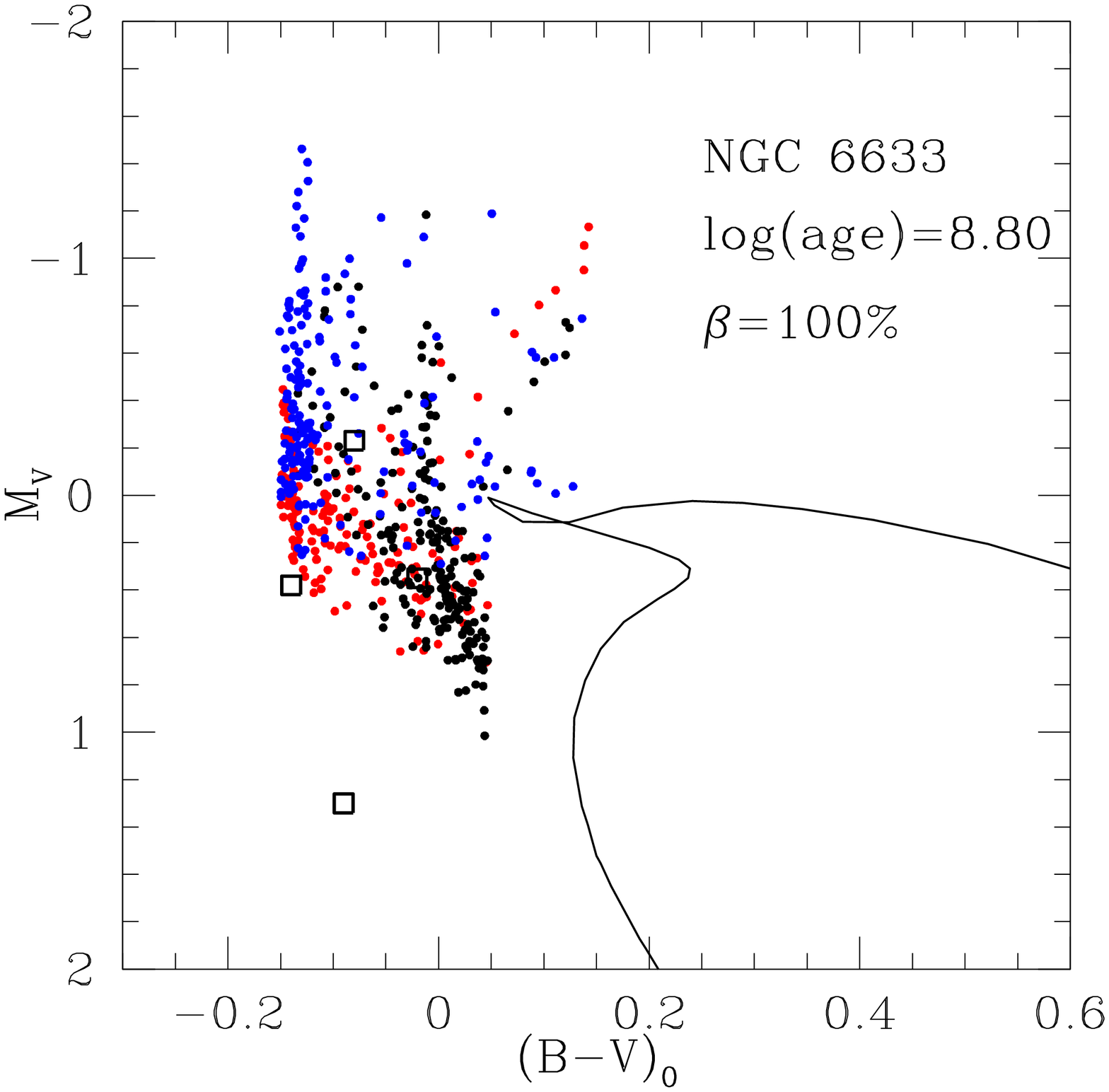}
\hfill
\includegraphics[scale=0.24]{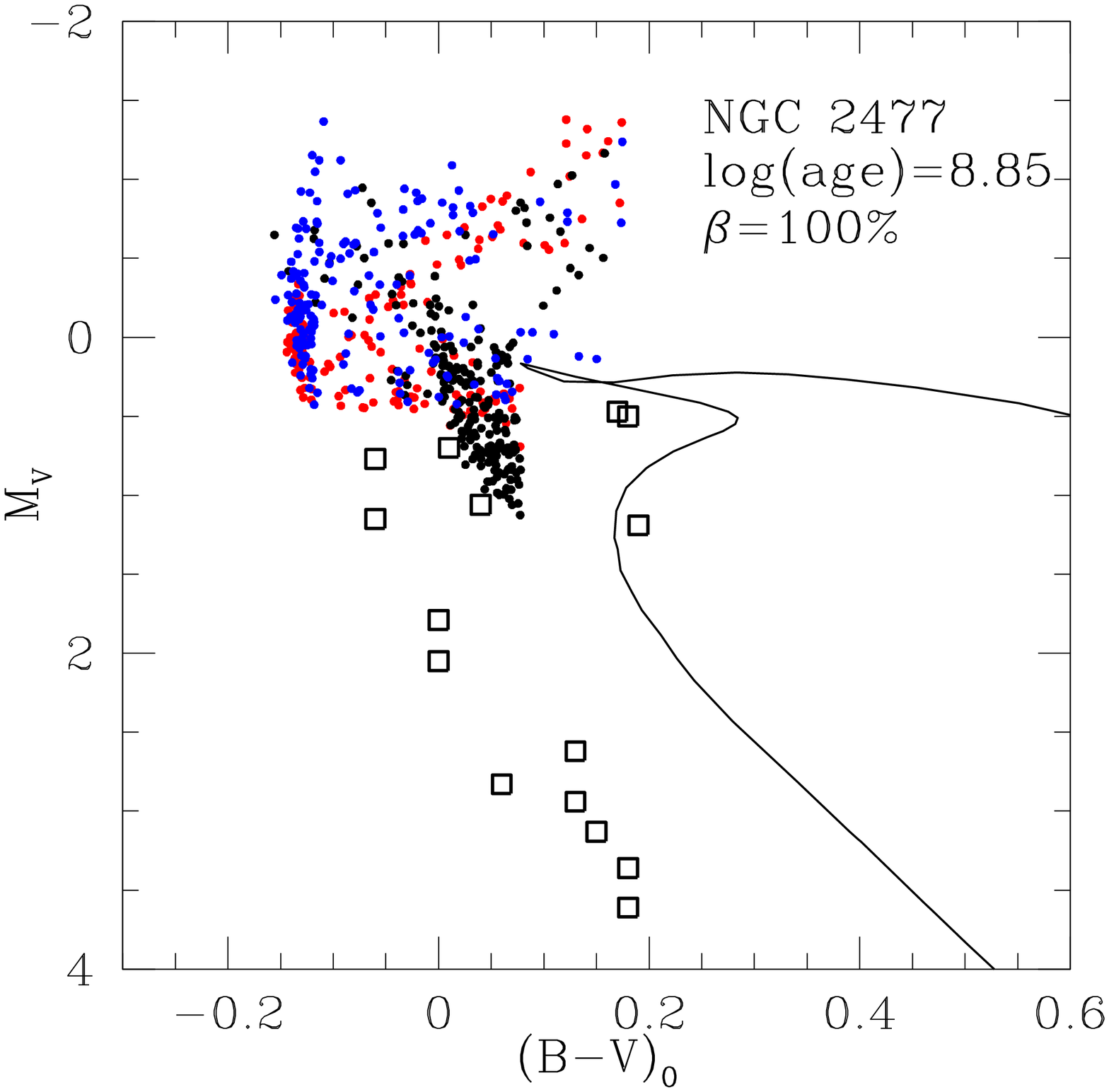}
\hfill
\includegraphics[scale=0.24]{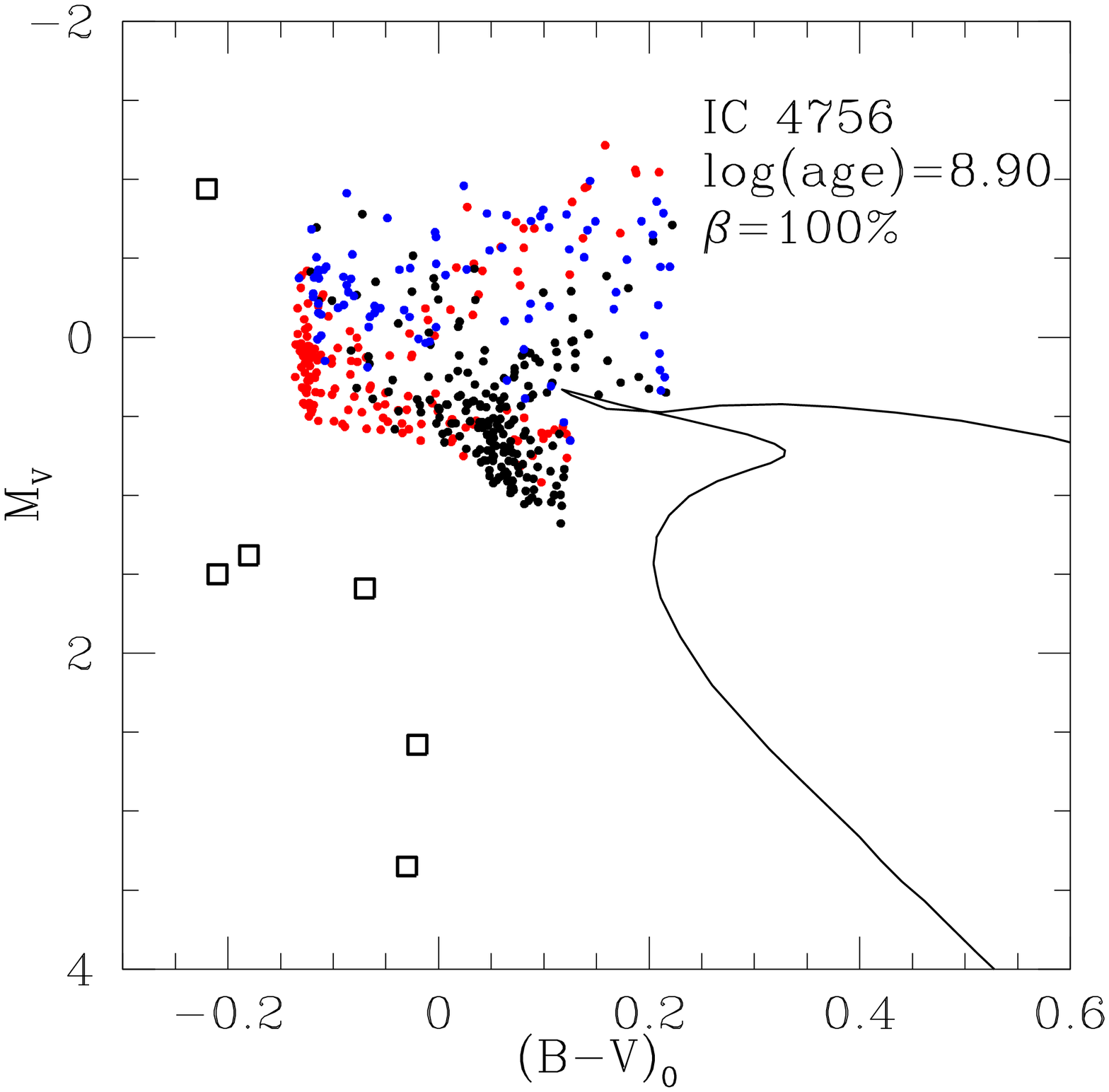}
\hfill
\includegraphics[scale=0.24]{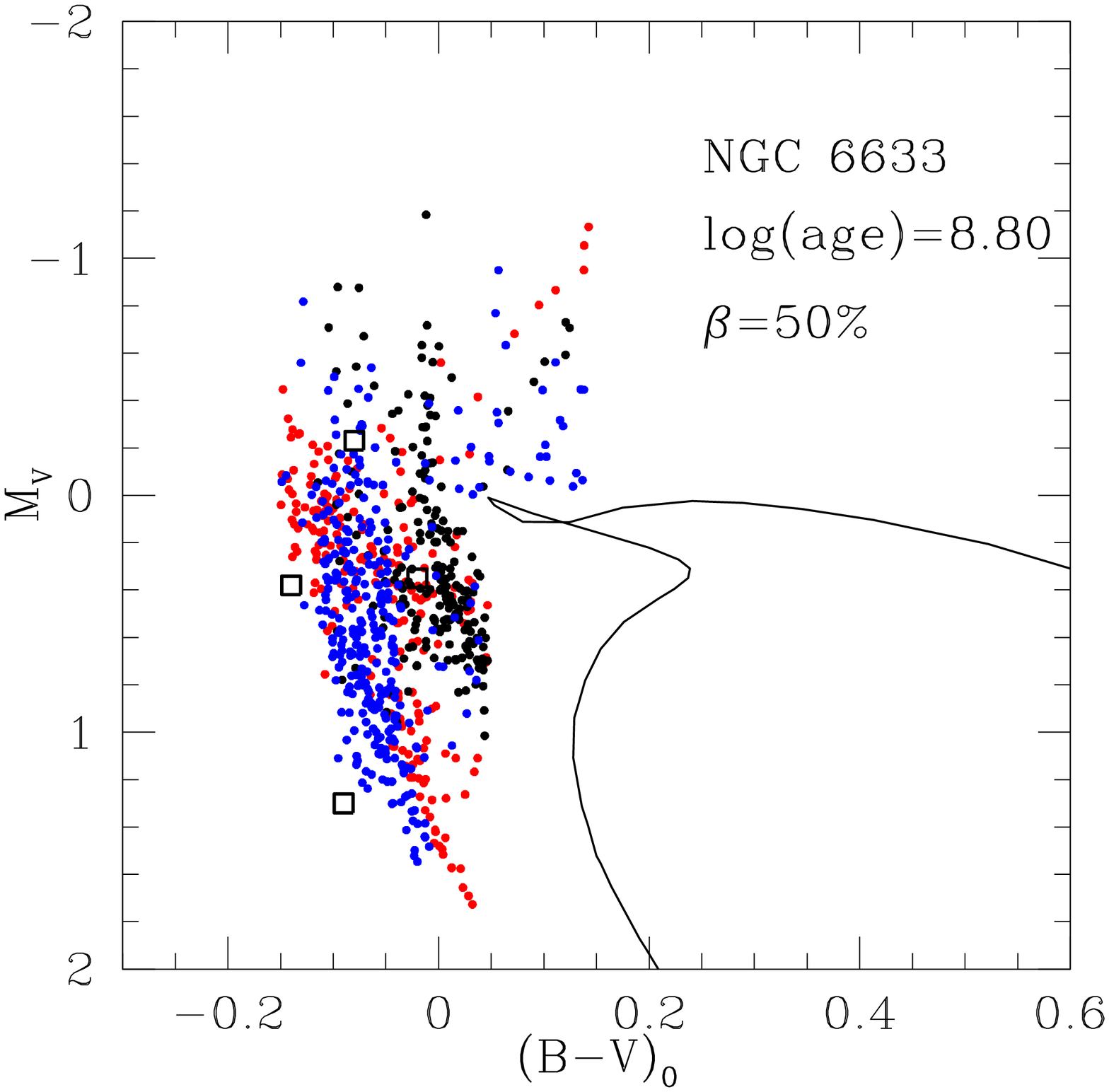}
\hfill
\includegraphics[scale=0.24]{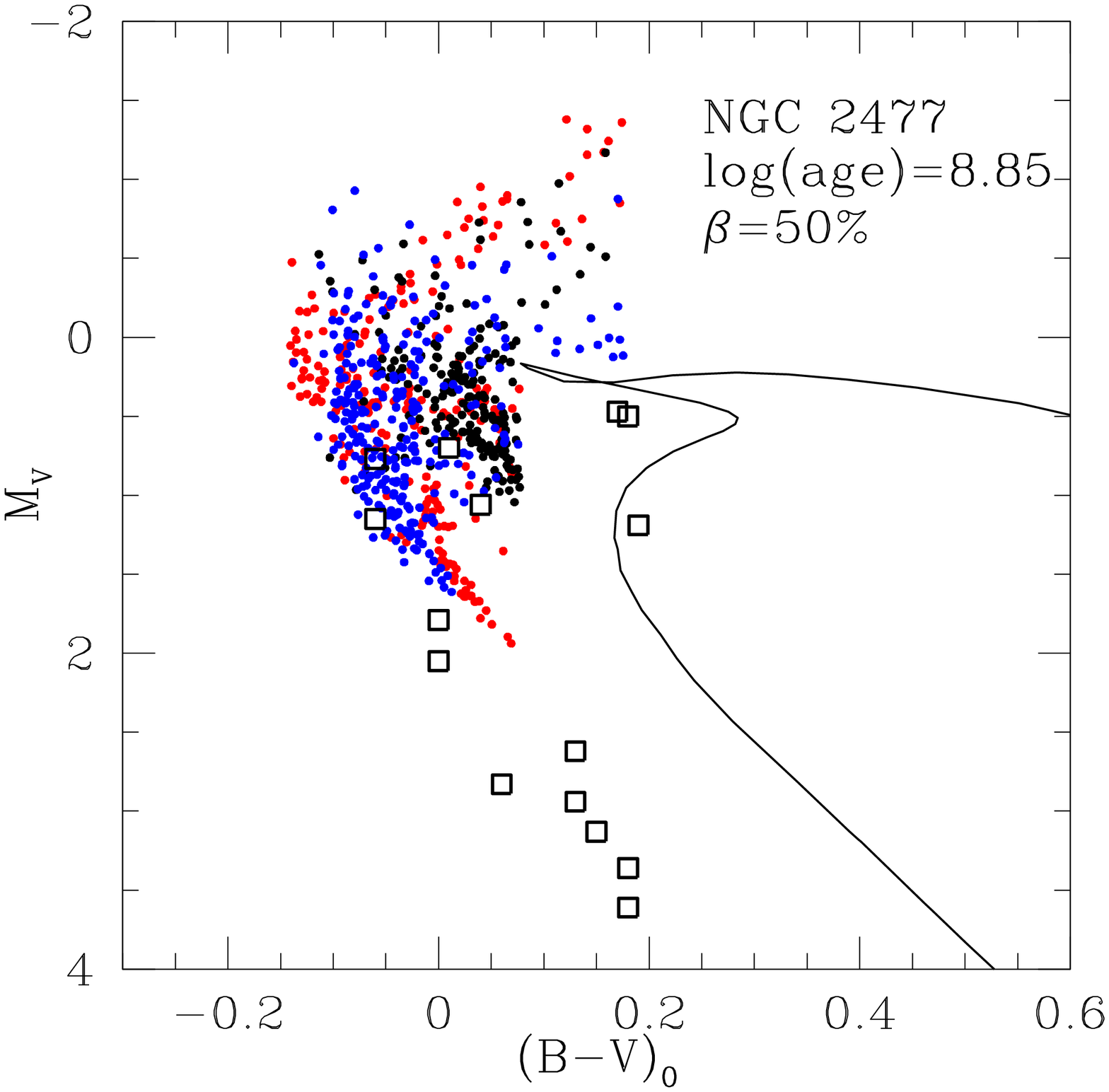}
\hfill
\includegraphics[scale=0.24]{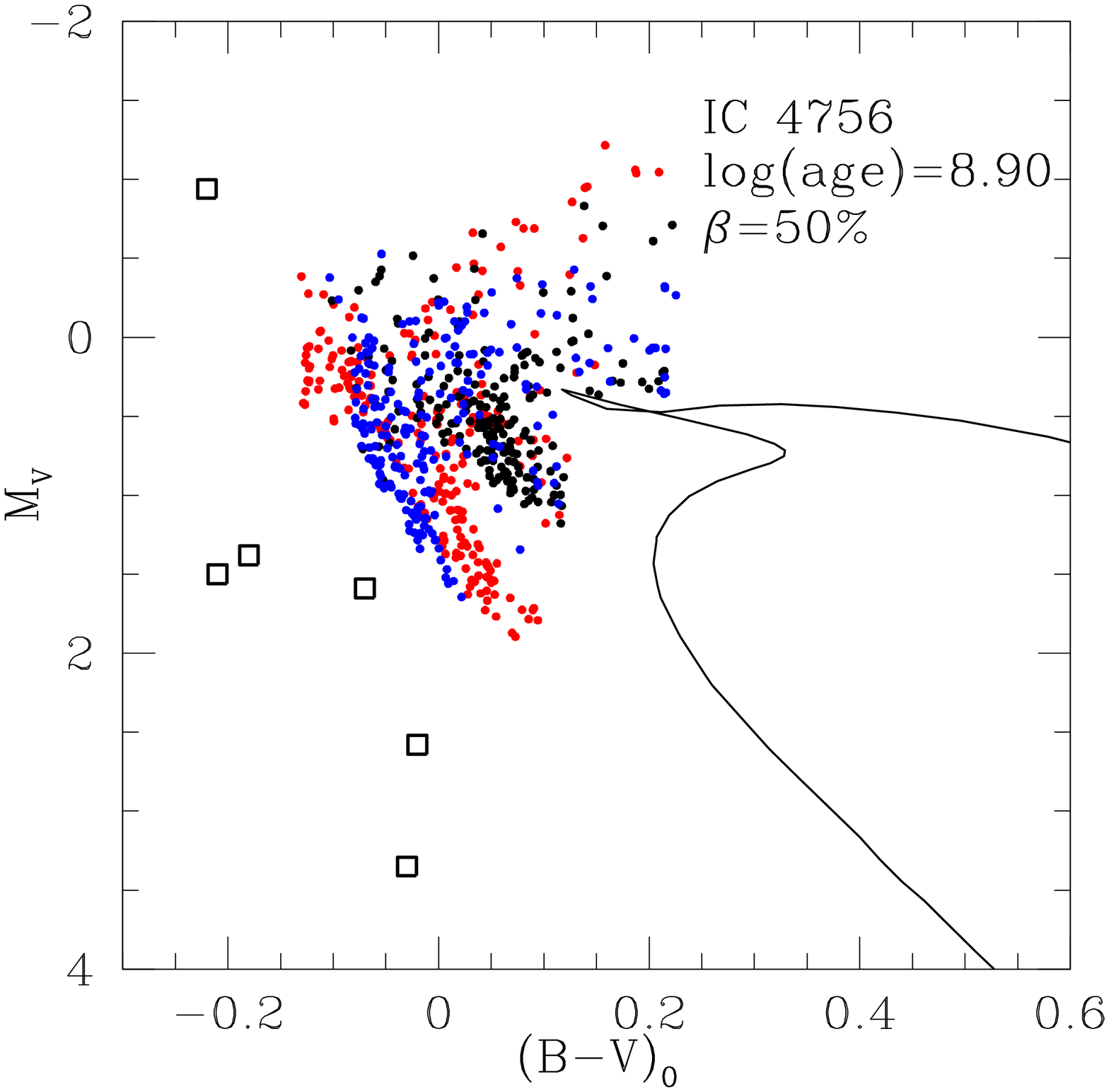}
\caption{CMDs of our simulated star clusters with $\beta$=1 on the
upper and $\beta$=0.5 on the lower panel for each pair. The samples
are IC 2488, IC 2714, NGC 1245, NGC 6633, NGC 2477 and IC 4756 in
order. Open squares are the observed BSs from sample clusters in
AL07. Points colored with black, blue and red are BSs formed via
case A, B and mergers respectively. \label{f:CMD}}
\end{figure*}

A sample of clusters in AL07 are selected based on the age and
metallicity in order to check our simulations. The fundamental
parameters of these selected clusters are listed in table 2. In order
to facilitate the comparison, firstly we need to pay attention to
the identification criterion of BSs in AL07. A cluster star is
assumed to be a blue straggler candidate if it is located in the
shaded area in the CMD of the cluster as showed in Fig~\ref{f:def}.
In addition in the shaded area, it is also physically reasonable to
consider stars that has a blue offset of ZAMS and stars that are brighter
than and somewhat redder than the TO of the cluster. We adopt the
selection criterion similar to AL07 in our work. Simulations for the
BS populations in six clusters, namely IC 2488, Ic 2714, NGC 1245,
NGC 6633, NGC 2477 and IC 4756, are presented in Fig~\ref{f:CMD}. In
fact, realistic simulations of the sample clusters may have large
dispersion in the results due to the relative low number of massive
member stars in young clusters. Therefore, the simulations are made
with $10^6$ artificial primordial binaries so that the number of
theoretical BSs in Fig~\ref{f:CMD} is of course much larger than
observations, this is only considered as the indication of the
possible locations of the BS models. The results are shown in
Fig~\ref{f:CMD} in which a pair of CMDs are given for each of the
selected clusters. The upper one is for the conservative case while
the lower one is for the non-conservative case. In each of the
plots, the observed BSs are shown in open squares. The solid line is
the corresponding isochrone that best fits the cluster parameters
(also indicated on the figure). Dots are simulated BS models. Points
colored with black, blue and red are BSs formed via case A, B and
mergers respectively.

The distribution of BSs on CMDs are different between the
conservative and the non-conservative cases. We have adopted a
constant $\beta$ in LDZ10 to estimate BSs formed via
non-conservative case in M67. The result shows in M67 a high value
of $\beta$ can reproduce the luminous BSs better, while lower value
can match the number of BSs with low luminosity better. Both results
show that the number of BSs could be significantly increased by
taking into account Case B.

In this work, we also adopt $\beta$=0.5 as many previous work for case B mass transfer
to simulate non-conservative cases.
This means that 50\% of the mass transferred during RLOF will be lost. After we considering this assumption, the most prominent
changes on CMD is that the luminosity of case B models decreases,
but the number of BS candidates increases. We can see from
Fig~\ref{f:CMD} that in some of the clusters, the lower mass
transfer efficiency is needed to fit the observed BSs, such as NGC
6633 which have 4 BS candidates (HD 170563, HD 170472, HD 169959 and
HD 170054). By taking $\beta$=0.5, the faintest BS (HD 170054) can
almost be covered by our models. HD 170563 might be a BS formed by
case A mass transfer according to its location on CMD. IC 2488 is a
young cluster with log(age)=8.25, in which the observed BS can be
easily confined by the conservative models. The two BSs to the blue
side are most likely formed via case B or mergers. In IC 2714, we
noticed that either the conservative or the non-conservative models
can fit the observational BSs because the two stars are both located
very close to the ZAMS. We also noticed that not all our models can
explain the observed BSs in all these clusters as an entity. All the
observed BSs in NGC 1245 are too faint compared to the cluster's
age. Although the BSs formed via AML can contribute to low
luminosity BSs, the cluster is so young that role of AML can not be
so remarkable. There is a clear shift in V band in NGC 2477. We can
not match the observations unless a shift of $\thicksim1.5$ mag in
distance modulus(DM) or we adopt an older cluster age. In fact, the cluster age is
different from 1.0 Gyr as from Xin et al. (2007), hereafter Xin07.
The parameters of this cluster needs to be confirmed. In IC 4756,
all the observed BSs show an extreme color toward the blue side and
an wide distribution in the magnitude. The color exceeds 0.2 toward
the blue side from the color limit of our simulation. Enhancement of
surface helium abundance of BSs may lead to a bluer color (Chen \&
Han 2004, 2008), but 0.2 is too big for any possible mechanisms.
The reason for He enrichment might be that they were formed via mass
transfer between an AGB and a MS star. However, an AGB star can only
provide He rich material if it is older than 2.24 Gyr for Population
I (Chen \& Han 2009), it is difficult to imagine that all of the BSs
in IC 4756 are formed in that single way. The observed BSs of IC
4756 can not be explained by mass transfer or merger in our work,
if the observed colors are correct.

Two distinct sequences of BS populations are found in M30 which can
be explained by direct collisions and evolution of close binaries
respectively. We notice two distinct sequences in our samples
similar to those in M30, such as NGC 1254, NGC 6633, IC 2477 and IC
4756 when $\beta$=1 (Fig~\ref{f:CMD}), however, we have different
explanations. These two sequences are mainly from different
evolutionary channels. Case B and merger models dominate the blue
sequence, in which case B mainly populate the luminous part of the
sequence. The case A models are only visible in a region close to
the ZAMS in relatively older clusters. When we take $\beta$=0.5, the
BS formed via case B has a luminosity about 1.0 mag lower and
degenerate with the mergers, or moves closer to the ZAMS (IC 2488),
so that the sequences can not be observed.

In some young clusters such as IC 2488 and IC 2714, we can notice
only one sequence located to the blue side of the ZAMS which is
formed by mergers and case B, the other sequence formed via case A
prominent in older clusters is not visible any more. We infer
from the results that case A can hardly contribute to the observed BSs in young
clusters as we concluded in our previous work (Lu \& Deng 2008). It
can also be explained by our models in three aspects.

(I)With a flat distribution of initial orbital separations in
$\log(\alpha)$, BS production due to case B models is dominant in
massive binaries.

(II)The durations of RLOF in case B models become shorter for higher
initial donor masses. The companions in massive binaries tend to
spend less time on RLOF and evolve faster to the blue side,
possessing most of the system masses. Then they evolve for the rest
of their lives as BSs.

(III)BSs formed via case A which still experiencing mass transfer
tend to form a redder BS sequence $\thicksim 0.75$ mag above the
ZAMS in BS region. Those BSs can sometimes degenerate with the MS
stars that are sitting on nearly vertical ZAMS in young clusters.

\subsection{The specific frequency}

In table 2, apart from the basic physical parameters, statistics of BS
populations in the sample clusters are also given. Column 7, 8, 9
are respectively $N_2$, $N_{\rm BS}$ and the corresponding specific
frequencies of BS defined as $f$=$N_{\rm BS}/N_2$. The value of $N_2$
and $N_{\rm BS}$ are significantly different among various authors. For
NGC 2516 and NGC 5617, the values of $N_{\rm BS}$ are quite different
between AL07 and Xin07. For NGC 2539, the value of $N_{\rm BS}$ agrees well
in AL07 and Xin07. However, the value of $N_2$ of AL07 is five times
that of Xin07. There are also some uncertainties in ages and
distance modulus from observations. For instance, the age of NGC
2477 is 8.85 in AL07 and 9.00 in Xin07.

\begin{figure}
 \centering
\includegraphics[scale=0.50]{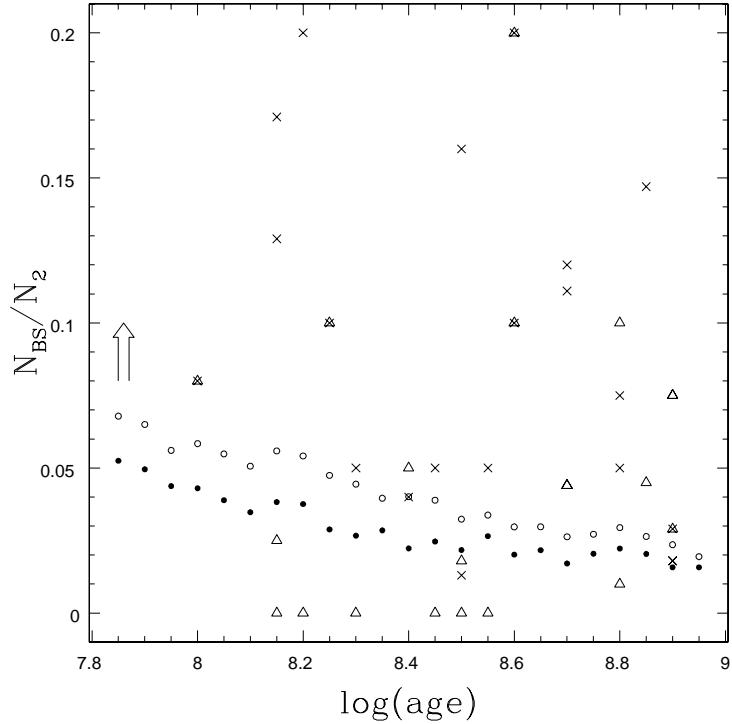}
\caption{Frequency $f$=$N_{\rm BS}/N_2$ versus cluster age. Dots
($\beta$=1.0) and open circles ($\beta$=0.5) are the result from our
simulation. Open triangles and crosses are $f$ from AL07 and Xin07
respectively.\label{f:bsn2}}
\end{figure}

It is believed that there might be a relationship between the number
of BSs and the total number of stars usually represented by stars
within 2 mag below the TO. Xin07 gave the $N_{\rm BS}$ distributions
with respect to three different parameters: $\log(age)$, $Z$, and
$N_2$ based on observations. The only correlation found is that
between $N_{\rm BS}$ and $N_2$. Chen \& Han (2009) discussed this issue
and estimated the frequency of $N_{\rm BS}$/$N_2$ with theoretical
models. The result showed that the frequency decreased with
increasing log(age) between 8.0 to 10.0. The AML becomes important
only when $\log$(age)$>$9.0. The contribution of BSs formed via
coalescence to the specific frequency is neglectable (Chen \& Han
2009). We also estimated the relationship of $N_{\rm BS}$/$N_2$ with
$\log(age)$ from 7.85 to 8.90 using our models. The results are
presented in Fig~\ref{f:bsn2} with solid dots. There seems to be a
correlation between age and $f$. Our results agree with Chen \& Han
(2009) but our $f$ are a little higher. The reason could be that we take
mergers due to low mass ratio binaries into account and use a
different definition of $N_2$. We use the same method as Hurley,
Tout \& Pols (2002) to treat a binary as a single star in our work
while in Chen \& Han (2009), a binary is assumed to be two single
stars and the evolution of primary has not been followed after RLOF.
$f$ from the non-conservative case are also plotted in
Fig~\ref{f:bsn2} with open circles. We can also notice that the
specific frequency is higher than the conservative case because the
number of BSs (usually fainter) will increase for lower $\beta$, as
discussed above. The observed frequencies in terms of $N_{\rm BS}/N_2$
for sample clusters from AL07 (open triangles) and Xin07 (crosses)
are also shown in the figure. We can see from the figure that only a
few clusters fall into the simulation predictions. Obviously, the
dispersion of observed specific frequency of BSs is too large as
discussed in the beginning of this section, primarily due to
stochastic effects of low member stars in these clusters. Dynamic
interactions between member stars within the cluster and with the
environments also play an important role in the scatter of the
observed $f$. Our simulation considers only stellar evolution
therefore can not fully interpret observations. More realistic
simulations for open clusters should take into account all these
effects, which is out of the scope of the current work.

There are also uncertainties in initial parameters of our
simulations (ie. IMF, distribution of $q$ and orbital separations).
The parameters may be different from reality. For instance, IMF is
still an open problem and can not be sure if it is universal in any
stellar systems or not. The binary frequency in star clusters is
also a very important issue for BS formation. Moreover, dynamical
processes in star clusters may also have influences though limited
in young clusters. The observational $N_{\rm BS}/N_2$ needs to be
normalized to take a better comparison.

After all, there are some difficulties in comparing $N_{\rm BS}$/$N_2$
in simulations and observations. There are very few BSs in young
clusters due to low number of member stars (few $N_2$ stars). The
$N_{\rm BS}-N_2$ correlation is still inchoative for currently available
data. More observations are needed to study the relations between
$N_{\rm BS}$ and $N_2$.

\section{Summary and conclusions}
\label{sect:sum}

We investigated BSs generated from mass transfer scenario and
simulated BS populations in young clusters in this work. BSs can be
produced via various mechanisms (Piotto et al. 2004). Observed BSs
in clusters can not be accounted for by a single mechanisms. In
general, BSs could be formed by mass transfer in primordial
binaries, which dominates BS formation in a spare environment
(Mathys 1991), or through direct collisions, which are considered to
be crucial in dense stellar environment (Fregeau et al. 2004).

The dynamical processes are not considered in this work because we
would like to discuss those BSs only formed in mass transfer
scenario. We choose young clusters as working samples in order to
avoid the complicated effects of dynamical evolution. In reality,
dynamic processes destroy primordial binaries while creating new
ones through capture and exchange processes (Hurley et al. 2005).
Mass segregation can also enhance BS formation in cluster cores.
Magnetic activities, as discussed in Sect 2, can also affect the
number of BSs through AML (Li et al. 2004; Demircan et al. 2006;
Micheal \& Kevin 2006; Stepien 2006), however it is only important
in clusters older than $10^9$ yr (Chen \& Han 2009).

We apply the same method as LDZ10 to calculate a grid of close
binary evolution models undergoing both case A and B RLOF in a wider
donor mass range. The formation and lifetime of BS are thoroughly
analyzed using the models. Our results show that case B produces
bluer, brighter and longer lived BSs than case A, therefore case B
is more effective in producing BSs than case A, especially in young
clusters.

Based on the grid of all our models at a given age, Monte-Carlo
simulations were carried out for a number of clusters of AL07. The
results show that BSs formed via different scenarios populate
different areas on CMD. Mass transfer efficiency is considered in
this work. A constant mass transfer efficiency of 0.5 is adopted for
the comparison between conservative and non-conservative cases. The
results show a high value of $\beta$ (conservative case) can
reproduce luminous BSs through case B scenario while lower value
case ($\beta$ = 0.5) can increase the number of BSs with low
luminosity. Case A is less important in young clusters as predicted
in Lu \& Deng (2008). Case B and merger become more important for
clusters with younger ages. The specific frequency of $f$ =
$N_{\rm BS}$/$N_2$ is studied. The result is less conclusive in the
current scheme.

Two separated sequences can be noticed in CMD in the clusters
between $\log(age)>8.7$ up to nearly 1.0 Gyr (Fig~\ref{f:bsn2}).
Case B and merger models dominate the blue sequence, in which case B
mainly populate the luminous part of the sequence. The case A models
are only visible in a region close to the ZAMS in relatively older
clusters. This may provide a new clue to the two sequences of BSs
observed in star globular clusters (Ferraro et al. 2009).

The grid of detailed binary evolution models in this work covers a
large parameter space, which can account for all possible binary
mass transfer processes. Therefore, this grid of models can serve
future studies on star clusters. By implementing dynamics, more
realistic simulation of star clusters, including under populated
open clusters, is possible. This will be part of our future work.

\section{Acknowledgments}
We are grateful to the referee for the useful and inspirational
comments that helped to improve this work. This work is supported by
the National Science Foundation of China through Grants No.
10973015, No. 10773015 \& No. 11061120454.

\label{lastpage}

\end{document}